\documentclass[10pt,journal,compsoc]{IEEEtran}
\IEEEoverridecommandlockouts
\usepackage{caption}
\usepackage{subcaption}
\usepackage{hyperref}
\hypersetup{
    colorlinks,
    citecolor=black,
    filecolor=black,
    linkcolor=black,
    urlcolor=black
}

\usepackage{array, tabularx, caption, boldline}
\usepackage[flushleft]{threeparttable}
\usepackage{cellspace}
\def\ps@headings{%
\def\@oddhead{\mbox{}\scriptsize\rightmark \hfil \thepage}%
\def\@evenhead{\scriptsize\thepage \hfil \leftmark\mbox{}}%
\def\@oddfoot{}%
\def\@evenfoot{}}
\usepackage{epsfig, epsf, amsmath, array, amsthm, amssymb, latexsym, graphics, multirow, cuted}
\usepackage{graphicx,color}
\usepackage{color}
\usepackage[ruled]{algorithm}
\usepackage{algorithmic}

\makeatletter
\usepackage{amsthm}
\usepackage{amsmath}
\newcommand {\mymarginpar}[1]{\bigskip\marginpar{#1}\bigskip}
\newcommand{\normalmode}{\renewcommand {\mymarginpar}[1]{}} 

\def\_{\rule{.3em}{.15ex}}      

\newcommand{\ls}[1]
   {\dimen0=\fontdimen6\the\font
    \lineskip=#1\dimen0
    \advance\lineskip.5\fontdimen5\the\font
    \advance\lineskip-\dimen0
    \lineskiplimit=.9\lineskip
    \baselineskip=\lineskip
    \advance\baselineskip\dimen0
    \normallineskip\lineskip
    \normallineskiplimit\lineskiplimit
    \normalbaselineskip\baselineskip
    \ignorespaces
   }

\newcommand {\bearn}{\begin{eqnarray*}}
\newcommand {\eearn}{\end{eqnarray*}}
\newcommand {\barr}{\begin{array}}
\newcommand {\earr}{\end{array}}

\newcommand {\N}{{\cal N}}

\newcommand\var[1]{\sigma_{#1}}

\def\defeq{\stackrel{\scriptstyle\rm def}{=}}
\def\twoLineSub#1#2{{#1}\atop{#2}}

\newtheorem{definition}{Definition}
\newtheorem{property}[definition]{Property}
\newtheorem{proposition}[definition]{Proposition}
\newtheorem{lemma}[definition]{Lemma}
\newtheorem{theorem}[definition]{Theorem}
\newtheorem{corollary}[definition]{Corollary}
\newtheorem{example}[definition]{Example}
\newtheorem{remark}[definition]{Remark}
\newtheorem{assumption}[definition]{Assumption}
\newtheorem{conjecture}[definition]{Conjecture}

\newcommand {\benum} {\begin{enumerate}}
\newcommand {\eenum} {\end{enumerate}}

\newcommand {\bdesc} {\begin{description}}
\newcommand {\edesc} {\end{description}}

\newcommand {\bfig}[2] {\begin{figure}[htbp]
                        \centerline {
                         \epsfig{figure={#1},clip=,width={#2}}}}
\newcommand {\bfigx}[2] {\begin{figure}[htbp]
						 \centerline {
						 \includegraphics[width=#2]{#1}}}                         
\newcommand {\brotatefig}[2] {\begin{figure}[htbp]
                        \centerline {
                         \epsfig{figure={#1},clip=,angle=-90,width={#2}}}}

\newcommand {\bfigfirst}[2] {\begin{figure}[h]
                        \centerline {
                        \setlength{\epsfxsize}{#2}
                        \epsffile{#1}}}
\newcommand {\efig}[2]{ \caption{#2}
                        \label{fig:#1}
                        \end{figure}
                        \mymarginpar{fig:#1}}
\newcommand {\erotatefig}[2]{ \caption{#2}
                        \label{fig:#1}
                        \end{figure}
                        \mymarginpar{fig:#1}}
\newcommand {\rfig}[1]{Figure \ref{fig:#1}}

\newcommand {\btab}[1]{
                       \begin{table}
                       \centering
                       \begin{tabular}{#1}}
\newcommand {\etab}[3] {
                       \end{tabular}
                       \caption[#3]{#2}
                       \label{tab:#1}
                       \end{table}
                       \mymarginpar{tab:#1}
                       \vspace{.1in}}

\newcommand {\btabular}[1]{\begin{center}
                       \begin{tabular}{#1}}
\newcommand {\etabular}{\end{tabular}
                       \end{center}}

\newcommand {\bdefin}[1]{\begin{definition}
                      \mymarginpar{def:#1}
                      \label{def:#1} }
\newcommand {\edefin}       {\end{definition}}

\newcommand {\bassum}[1]{\begin{assumption}
                      \mymarginpar{ass:#1}
                      \label{ass:#1} }
\newcommand {\eassum}       {\end{assumption}}
\newcommand {\rassum}[1]{Assumption \ref{ass:#1}}

\newcommand {\bpro}[1]{\begin{property}
                      \mymarginpar{pro:#1}
                      \label{pro:#1} }
\newcommand {\epro}   {\end{property}}

\newcommand {\bprop}[1]{\begin{proposition}
                      \mymarginpar{prop:#1}
                      \label{prop:#1} }
\newcommand {\eprop}       {\end{proposition}}

\newcommand {\blem}[1]{\begin{lemma}
                      \mymarginpar{lem:#1}
                      \label{lem:#1} }
\newcommand {\elem}   {\end{lemma}}
\newcommand {\rlem}[1]{Lemma \ref{lem:#1}}

\newcommand {\bthe}[1]{\begin{theorem}
                      \mymarginpar{the:#1}
                      \label{the:#1} }
\newcommand {\ethe}   {\end{theorem}}
\newcommand {\rthe}[1]{Theorem \ref{the:#1}}

\newcommand {\bconj}[1]{\begin{conjecture}
                      \mymarginpar{conj:#1}
                      \label{conj:#1} }
\newcommand {\econj}   {\end{conjecture}}

\newcommand {\bproof}[1]{\noindent {\bf Proof #1} \ }
\newcommand {\eproof} {\hfill \squares \\ \vspace{.3cm}}
\newcommand {\bcor}[1]{\begin{corollary}
                      \mymarginpar{cor:#1}
                      \label{cor:#1} }
\newcommand {\ecor}   {\end{corollary}}

\newcommand {\bax}[1]{\begin{axiom}
                      \mymarginpar{ax:#1}
                      \label{ax:#1} }
\newcommand {\eax}       {\vspace{-.1in} \end{axiom}}

\newcommand {\bex}[2]{\vspace{.1in}
                      \begin{example}
                      \mymarginpar{ex:#1}
                       {\bf #2}
                      \label{ex:#1} \em}
\newcommand {\eex}       {\end{example} \vspace{.3cm} }

\newcommand {\brem}[1]{\begin{remark}
                      \mymarginpar{rem:#1}
                      \label{rem:#1} \em }
\newcommand {\erem}   {\end{remark}}

\newcommand {\beq}[1]{\mymarginpar{eq:#1}
                      \begin{equation}
                      \label{eq:#1} }

\newcommand {\beqno}[1]{\mymarginpar{eq:#1}
                      \begin{eqnarray}
                      \nonumber}

\newcommand {\eeq}       {\end{equation}}
\newcommand {\eeqno}       { && \end{eqnarray}}
\newcommand {\req}[1]{(\ref{eq:#1})}

\newcommand {\bear}[1]{\mymarginpar{eq:#1}
                       \begin{eqnarray}
                       \label{eq:#1} }

\newcommand {\bearno}[1]{\mymarginpar{eq:#1}
                       \begin{eqnarray}
                       \nonumber}

\newcommand {\eear}{\end{eqnarray}}
\newcommand {\eearno}{\end{eqnarray}}
\newcommand {\bsel}{\left \{ \begin{array}{cl}}
\newcommand {\esel}{\end{array} \right.}

\newcommand {\bmat}[1]{\left [ \begin{array}{#1}}
\newcommand {\emat}{\end{array} \right ]}

\def\R{I\kern-0.30em R}
\def\N{I\kern-0.30em N}
\def\P{I\kern-0.30em P}

\newcommand\squares{\vrule height6pt width7pt depth1pt}

\def\ex{{\bf\sf E}}
\def\pr{{\bf\sf P}}
\def\cov{{\bf\sf Cov}}

\def\twoLineSub#1#2{{#1}\atop{#2}}

\newcommand{\binomcoeff}[2]{\left(\begin{array}{c}#1\\#2\end{array}\right)}

\def\etal{{\em et al.}\ }

\normalmode

\title{ A generalized configuration model with triadic closure}

\author{Ruhui~Zhang,
        Duan-Shin~Lee,~\IEEEmembership{Member,~IEEE,}
        and Cheng-Shang~Chang,~\IEEEmembership{Fellow,~IEEE}
\IEEEcompsocitemizethanks{\IEEEcompsocthanksitem Ruhui Zhang is with the Department
of Computer Science, and Duan-Shin~Lee
is with the Department of Computer Science and the Institute of Communications Engineering, and 
Cheng-Shang Chang is with the Institute of Communications Engineering,  
National Tsing Hua University, Hsinchu 300, Taiwan, R.O.C. (Email:huibrana@gapp.nthu.edu.tw; lds@cs.nthu.edu.tw;cschang@ee.nthu.edu.tw)}
\thanks{This research was supported in part by the
   Ministry of Science and Technology,Taiwan, R.O.C., under Contract 109-2221-E-007-093-MY2.}}

\begin{document}
\IEEEtitleabstractindextext{
\begin{abstract}
In this paper we present a generalized configuration model with random triadic closure
(GCTC). This model possesses five fundamental properties: large clustering coefficient, 
power law degree distribution, short path length, non-zero Pearson 
degree correlation, and existence
of community structures.  We analytically derive the
Pearson degree correlation coefficient and the clustering coefficient of the proposed model. 
We select a few datasets of real-world networks.  By simulation, we show that
the GCTC model matches very well with the datasets in terms of Pearson
degree correlations and clustering coefficients.  We also test three well-known community detection algorithms
on our model, the datasets and other three prevalent benchmark models. 
We show that the GCTC model performs equally well as the other three benchmark models.
Finally, we perform influence diffusion on the GCTC model using the independent 
cascade model and the linear threshold model.  We show that the influence spreads of
the GCTC model are much closer to those of the datasets than the other benchmark
models.  This suggests that the GCTC model is a suitable tool to study network
science problems where degree correlation or clustering plays an important
role.
\end{abstract}


}

\maketitle

\IEEEraisesectionheading{\section{Introduction}\label{s:introduction}}
\IEEEPARstart{N}{etwork} science is an inter-disciplinary field that formulates research problems arising
in science and engineering into problems in graphs.  To study successfully these 
diversified problems,
researchers need suitable graph models.  It would be nice to 
have a random graph model that possesses a rich set of topological properties and
can be used to study a wide variety of problems in network science.  In the mean time, 
it would be nice that the random graph model is simple enough to allow mathematical
studies.  There are five properties that are commonly observed in many
real-life networks in science and engineering \cite{Newman2010, Toivonen2006}. They are
\begin{enumerate}
\item large clustering coefficient;
\item short average path length;
\item scale-free degree distribution, {\em i.e.} power law distribution; 
\item assortative or disassortative degree correlation, and
\item existence of community structures.
\end{enumerate}

In this paper we propose a random network model that possesses the five properties above.
Many research problems in network science tie closely with these five properties.  For instance, it is well known that diseases transmit efficiently among people
in densely connected clusters, such as members in a household 
\cite{Coupechoux.2014,trapman2007analytical}. It is also well known that clustering
affects significantly the resilience of a network \cite{gleeson2010clustering}.
It is also commonly observed that densely connected clusters affect significantly how
opinions spread in a network \cite{morris2000,Watts.2007}. To study these problems, one
needs a graph model that is rich in transitivity.
Authors in \cite{Boguna2003,Boguna2002,Eguiluz2002} showed that degree correlations
and power law degree distributions have strong influence to the threshold of an epidemic.
To study epidemic or influence diffusion problems, one may need a graph model that possesses
degree correlations.
The purpose of this paper is to propose a random graph model that is suitable
to be used to study a broad range of problems.  Our random model has all the five properties
listed in the previous paragraph and is simple enough to allow mathematical analysis.  
It has a rich
set of parameters and allows users to match its degree distribution, clustering 
coefficient, and degree correlation with those of a real-world dataset.

Our proposed model is based on configuration models originally proposed by
Bender \cite{Bender1978}. A configuration model is defined
and constructed for a given degree sequence.  It can have a power law degree distribution,
if the given degree sequence is drawn from a power law distribution.
It possesses a small world property.  However, it lacks a community structure, and its
clustering coefficient and degree correlation are asymptotically small as the network grows
in size.  Lee \etal \cite{Lee2019} proposed a generalized configuration model, which adds a positive or 
a negative degree correlation to the configuration model.
Lee {\em et al.} \cite{Lee2019} achieved a non-zero degree correlation by partitioning stubs of 
edges into blocks and connecting stubs according to certain rules.  
In this paper, we propose to add another layer of blocks to emulate an artificial
structure of communities.  The result is a random graph model that possesses the
fifth property listed in the first paragraph of this section.
In addition, we propose to add triangles into the model to create a significant
clustering coefficient.  In literature, there are several proposals to 
achieve this objective.  We refer the reader to \cite{Newman2009,Wang2014} and
the references therein.
In this paper we propose to add triangles by performing triadic closure 
operations \cite{Rapoport1953, Easley2010}.  In social science,
triadic closure means that there is an increased likelihood that two people, 
who have a friend in common, will become friends \cite{Rapoport1953}. Triadic 
closure has been observed in many real-world networks. For instance, 
Kossinets and Watts found clear evidence of triadic closure by taking multiple 
snapshots on an email communication network using a dataset consisting of 
22,000 students at a large U.S. university \cite{Kossinets2006empirical}. 
Moreover, Leskovec \etal \cite{Leskovec2008microscopic} analyzed the 
properties of triadic closure in online social networks of LinkedIn, Flickr, 
Del.icio.us, and Yahoo! Answers.  In this paper we propose a random triadic
closure operation.  That is, all connected triples that are not triangles yet
are closed and become triangles with some probabilities. 
With these two new features, the new model possesses large transitivity and community
structures.  We call the new model generalized configuration model with triadic 
closure (GCTC).   The detail construction algorithm of the GCTC model will be presented in 
Section \ref{s:model}. We mention some research work
in the literature that is related to triadic closure.  Zhou et al. \cite{Zhou_2018}
proposed a network embedding method that takes the status of triads into consideration.
Their method can learn representation vector for each vertex at different time points.
Hofstad et al. \cite{van_der_Hofstad_2018} used term ``triadic closure" to refer to
triangles in a network.  Hofstad et al. studied configuration models with power law
degree distributions.  Specifically, they showed that the local clustering coefficient
of vertices with degree $k$ and network size $n$ is of the order $n^{5-2\tau}k^{-2(3-\tau)}$
for $\tau\in(2, 3)$, where $\tau$ is the exponent of the power law degree distribution
\cite{van_der_Hofstad_2018}.  Clustering coefficients have been a widely accepted 
metric to measure triadic closure.  Yin et al. \cite{Yin_2020} proposed new
metrics to measure triadic closure in directed graphs.

We now outline the contributions of this paper.  First, we derive closed form expressions for the 
Pearson degree correlation coefficient and the clustering coefficient of the 
new proposed GCTC model.
We mention that there are other proposals of random networks that possess the five 
fundamental properties listed above.  For example,
Toivonen \etal \cite{Toivonen2006} proposed a growth model that possesses 
all the five properties. It is not clear if the 
Pearson degree correlation coefficient of this model is mathematically tractable.
Second, we examine whether the GCTC network is a suitable model to study various
research problems in network science.  We choose four datasets of networks collected
in the real world.  To make a comparison, we choose
three random network models that are often used as benchmark models to evaluate
community detection algorithms.  The three models are stochastic 
blockmodel (SBM) \cite{Holland1983, Karrer2011}, 
Lancichinetti-Fortunato-Radicchi benchmark (LFR) model
\cite{Lancichinetti2008benchmark, Lancichinetti2009benchmarks} and 
artificial benchmark for community detection (ABCD) model \cite{Kami_2021}.  
Through simulations, we show that by choosing parameters properly 
the GCTC model can match better with the datasets in terms of the Pearson degree
correlations and the clustering coefficients than the SBM, the LFR model and the ABCD model.  
Further, we compare the GCTC model with the SBM, the LFR model and the ABCD model using three well 
known community detection algorithms.  We show that
the GCTC model performs equally well as the other three benchmark models for community detection.
Finally, we study influence diffusion in real-world networks, the GCTC model, the SBM, LFR and ABCD models.
We study this problem using both an independent cascade (IC) model 
\cite{kempe2003,shakarian2015independent} and a linear threshold (LT) model 
\cite{kempe2003,borodin2010threshold}.  We find that the fraction
of influenced nodes observed in the GCTC model is much closer to 
that of the real-world datasets, compared with the fraction corresponding to the SBM model, LFR,
and the ABCD models.  This implies that the GCTC network is a more
suitable model to study the influence diffusion problem.  These results
seem to indicate that degree correlation and clustering coefficient are less relevant to
the community detection problem, but are crucial to the influence diffusion problem.

The organization of this paper is as follows.
In Section \ref{s:model}, we present a construction algorithm for the
GCTC model. Then, we briefly review several basic analytical results of the generalized
configuration model in Section \ref{s:review}. In Section \ref{s:pdcc} we 
analyze the Pearson degree correlation coefficient of the GCTC model.
In Section \ref{s:lcc} we analyze the clustering coefficient of the GCTC model.
In Section \ref{s:numerical} we present numerical and simulation
results.  Finally, we present the conclusions of this paper in 
Section \ref{s:conclusions}.

\section{Construction Algorithm}\label{s:model}

In this section we present a construction algorithm for the GCTC model.   Recall that in  a standard 
configuration model, two ends of an edge are called ``stubs".  
To construct a standard configuration model, an unconnected stub is connected to 
another randomly selected stub among all unconnected
stubs  \cite{Bollobas1980}.  This construction algorithm creates a random network 
that has asymptotically 
vanishing Pearson degree correlation coefficient as the network becomes large.
To introduce non-zero Pearson degree correlation, Lee \etal \cite{Lee2019}
partition stubs into blocks according to vertex degrees.  
To introduce positive
(resp. negative) correlation, the selected permutation function associates blocks of large 
(resp. small) degrees with another block of large degrees.  
Stubs in block $i$ are designated into type 1 stubs and type 2 stubs.  An unconnected 
type 1 stub in block $i$  is connected to a randomly selected unconnected type 1 stub
in the associated block of block $i$.  An unconnected type 2 stub is connected to an
unconnected type 2 stub randomly selected in {\em all} blocks. 
The construction algorithm of GCTC model is similar to that of 
a generalized configuration model, except that it has {\em two} additional features.  
First, GCTC model has two layers of blocks.  Stubs are divided into
macroscopic blocks, which model communities.  In each macroscopic block, stubs are
further divided into microscopic blocks, which are used to create a non-zero Pearson 
degree correlation.  To achieve this, each stub is designated to one of {\em three} types.
Type 1 and type 2 stubs provide intra-community connections and the non-zero degree correlation.
Type 3 stubs provide intra-community as well as inter-community connections.  

We describe the construction algorithm of GCTC model in details.
There are $c$ communities, where $c\ge 1$. Community $i$, where $i=1, 2, \ldots, c$, has
$n_i$ vertices.  Let $n$ be the total number of vertices in the network.  It follows
that 
\[
n=\sum_{i=1}^c n_i.
\]
We note that the size of different communities can be distinct.  That is, it is possible
that $n_i\ne n_j$ for some $i\ne j$. Denote $k_{ij}$ as the degree of the $j$-th vertex in community $i$.
Define $m_i$ such that
\beq{eqm_i}
2 m_i=\sum_{j=1}^{n_i} k_{ij}.
\eeq
The quantity $2m_i$ in \req{eqm_i} is the total number of stubs attached to vertices in community
$i$.  The total number of edges in
the network is $m$, where
\[
m=\sum_{i=1}^c m_i.
\]

For each community $i$, the edges are connected in the following ways. A degree $k$ vertex in community $i$ has $k_{i}$ stubs. 
In this community $i$, we arrange the stubs associated with the vertices in
an ascending order of the vertex degrees.  We then partition the
stubs into $b_i$ blocks evenly. That is, each block has the same number
of stubs.  Denote block $b_{ij}$ as the block $j$ in community $i$.
To create degree-degree correlations, we select a permutation function $h_{i}(b_{ij})$ for block $b_{ij}$.  Specifically, block $b_{ij}$ is associated with block
$b_{ii}$, if $h_i(b_{ij})=b_{ii}$. In addition, $h_i$ is selected such that $h_i(h_i(b_{ij}))=b_{ij}$.  We then classify the stubs in each block into three types proportionally.  
Denote the ratio of type 1 and type 2 stubs to the total stubs in each block  in community $i$ as $r_i\in[0, 1]$
and the ratio of type 1 stubs to the type 1 and type 2 stubs in each block  in community $i$ as $q_i\in [0, 1] $.
Suppose  $r_i$ and $q_i$ are given. For block $b_{ij}$, randomly designate 
$\lceil 2m_i q_ir_i/b_i\rceil$ stubs as type 1 stubs, and
randomly designate $\lceil 2m_i(1-q_i)r_i/b_i\rceil$ stubs as type 2 stubs.
Designate the rest stubs in block $b_{ij}$ as type 3 stubs. 
To make a connection, one randomly picks an unconnected stub, say stub $s$.
If $s$ is a type 1 stub in block $b_{ij}$ of community $i$, connect it
with a randomly selected unconnected type 1 stub in block $h_{i}(b_{ij})$ and connect it to $s$.
If $s$ is a type 2 stub, randomly select an unconnected type 2 stub in community
$i$ and connect it to $s$.  If $s$ is a type 3 stub, randomly
select an unconnected type 3 stub in the network and connect the stub to $s$.
These edges are referred to as {\em regular edges}.  

Next, we apply
triadic closure operations to increase the number of triangles
in the network.  The edges added into the network by the triadic
closure operations are called {\em transitive edges}.
We examine all pairs of unconnected vertices in the network.
For each pair of unconnected vertices, say vertices $A$ and $B$,
if $A$ and $B$ have $d$ 
common neighbors, we connect $A$ and $B$ 
with probability $a_d$. With probability $1-a_d$, $A$ and $B$ remain unconnected.
The construction algorithm for the GCTC model is shown in Algorithm \ref{alg1}.

\begin{algorithm}
	\caption{Construction Algorithm}\label{alg1}
	{\bf Inputs}: Degree sequence $\{k_{ij}:i=1, 2, \ldots, c, 
j=1, 2, \ldots, n_i\}$, parameters $\{b_i:i=1, 2, \ldots, c\}, \{q_i: i=1, 2, \ldots, c\}, 
\{r_i:i=1, 2, \ldots, c\}, \{h_i:i=1, 2, \ldots, c\}$ and $\{a_d: d=1,2,\ldots, n-2\}$.\newline
	{\bf Outputs}: graph $G=(V,E)$\newline
	\begin{algorithmic}[1]
		\FOR{$i=1, 2, \ldots, c$}
		\STATE For community $i$, create $2m_i$ stubs from degree sequence
		$\{k_{ij}: j=1, 2, \ldots, n_i\}$ and arrange the stubs
		in an ascending order according to the degrees;
		\STATE Divide $2m_i$ stubs into $b_i$ blocks evenly;
		\FOR{$j=1,2,\ldots, b_i$}
		\STATE For block $j$ of community $i$, randomly designate 
		$\lceil 2m_i q_ir_i/b_i\rceil$ stubs as type 1 stubs, and
		randomly designate $\lceil 2m_i(1-q_i)r_i/b_i\rceil$ stubs as type 2 stubs;
		\STATE Designate rest stubs in block $j$ as type 3 stubs;
		\ENDFOR
		\ENDFOR
		\WHILE{there are unconnected stubs}
		\STATE Randomly select a stub.  Assume that the stub is in block $j$ of
		community $i$;
		\IF{type 1 stub}
		\STATE connect this stub with a randomly selected type 1 
		 unconnected stub in block $h_i(j)$ in community $i$;
		\ELSIF{type 2 stub}
		\STATE connect this stub with a randomly selected type 2 
		unconnected stub in community $i$;
		\ELSE
		\STATE connect this stub with a randomly selected type 3 stub
		among all type 3 stubs that are unconnected in the network;
		\ENDIF
		\ENDWHILE
		\FOR{each unconnected pair of vertices}
		\IF{these two vertices have $d$  common neighbors}
		\STATE with probability $a_d$, these two vertices are connected with a transitive
		edge, and with probability $1-a_d$ leave these two vertices unconnected;
		\ENDIF
		\ENDFOR
	\end{algorithmic}
\end{algorithm}

\section{Review of the Generalized Configuration Model}\label{s:review}

In this section we review some basic results of the generalized configuration model
in \cite{Lee2019}.  These results will be used to derive Pearson degree correlation coefficient
of the GCTC model in Section \ref{s:pdcc} and the clustering coefficient of the GCTC model
in Section \ref{s:lcc}. We note that the generalized configuration model is a special case
of the GCTC model where $c=1, r_1=1$ and $a_d=0$. We drop subscripts
and use notation $b$, $q$, and $h(i)$ to denote the number of blocks, the
fraction of type 1 stubs, and the permutation function of block $i$, respectively. 

Let $Z$ be the degree of a randomly selected vertex in the generalized configuration network.
Then, the pmf of $Z$ is $\{p_k\}$ and
\beq{def-Z}
\ex[Z]=\sum_{k=0}^\infty k p_k.
\eeq
The following assumption is crucial to the analysis. This assumption is stringent.  
We refer readers to \cite{Lee2019} for additional information on this assumption.
\bassum{1}
The degree distribution $\{p_k\}$ is said to satisfy this assumption if
one can find mutually disjoint sets $H_1, H_2, \ldots, H_{b}$, such 
that
\[
\bigcup_{i=1}^b H_i=\{0, 1, 2, \ldots\}
\]
and
\beq{evenblock}
\sum_{k\in H_{i}} k p_k =\ex[Z]/b
\eeq
for all $i=1, 2, \ldots, b$.  In addition, we assume that the degree sequence
$k_1, k_2, \ldots, k_n$ sampled from the distribution $\{p_k\}$ can be evenly
placed in $b$ blocks.  That is, each block has the same number of stubs.
In addition, stubs that belong to vertices of the same degrees are placed
in the same block.
Mathematically speaking, there exist mutually disjoint sets $H_1, H_2, \ldots, H_{b}$
that satisfy
\begin{enumerate}
\item $\bigcup_{i=1}^b H_i=\{1, 2, \ldots, n\}$,
\item $k_i\ne k_j$ for any $i\in H_{\ell_1}$, $j\in H_{\ell_2}$, $\ell_1\ne\ell_2$, and
\item $\sum_{j\in H_i} k_j=2m/b$ for all $i=1, 2, \ldots, b$.
\end{enumerate}
\eassum

We randomly select a stub in the range $[1, 2m]$.  Denote this stub
by $t$.  Let $v$ be the vertex, with which stub $t$ is associated.  Let $Y$ be
the degree of vertex $v$. Since the stub
is randomly selected and vertices with degree $y$ have $n y p_y$ stubs. We have
\begin{align}
\Pr(Y=y)=\frac{n y p_y}{2m}
= \frac{y p_y}{\ex[Z]}.\label{marginal-pmf}
\end{align}

Now connect
stub $t$ to a randomly selected stub. 
Let this stub be denoted by $s$.  Let $u$ be
the vertex, with which $s$ is associated, and let $X$ be the degree of vertex $u$. Next we study $\pr(X=x | Y=y)$.
We assume that \rassum{1} holds. Suppose $x$ is a degree in set $H_i$. 
The total number of stubs which are associated with
vertices with degree $x$ is $n x p_x$.  By \rassum{1},
all $nx p_x$ stubs are in block $i$.  There are two cases, in which
stub $t$ connects to stub $s$.  In the first case, stub $t$ is of type 1.  This
occurs with probability $q$.  In this case, stub $s$ must be a type 1 stub and belong to a vertex
with a degree in block $h(i)$.  With probability
\beq{q}
\frac{qnxp_x}{2mq/b-\delta_{i, h(i)}},
\eeq
the  construction algorithm in Section \ref{s:model} connects $t$ to stub $s$.
In \req{q} $\delta_{i,j}$ is the Kronecker 
delta, is equal to one
if $i=j$, and is equal to zero otherwise. 
In the second case, stub $t$ is of type 2.
This occurs with probability $1-q$.  In this case, stub $s$ can be associated with a degree
in any block.
With probability 
\beq{1-q}
\frac{(1-q)nx p_x}{2m(1-q)-1}
\eeq
the construction algorithm connects stub $t$ to stub $s$.
Combining the two cases in \req{q} and \req{1-q}, we have
\beq{condp1}
\Pr(X=x | Y=y)
=\frac{q^2 nxp_x}{2mq/b-\delta_{i, h(i)}}+
\frac{(1-q)^2 nx p_x}{2m(1-q)-1}
\eeq
for $y\in H_{h(i)}$. 
If $y\in H_j$ for $j\ne h(i)$,
\beq{condp2}
\Pr(X=x | Y=y)= \frac{(1-q)^2 nxp_x}{2m(1-q)-1}.
\eeq
Now assume that the network is large.  That is, we
consider a sequence of constructed graphs, in which 
$n\to\infty$, $m\to\infty$, while keeping
$2m/n=\ex[Z]$.  
Under this asymptotic, Eqs. \req{condp1} and \req{condp2}
converge to
\begin{align}
\Pr(X=x | Y=y)\to
\left\{\begin{array}{ll}
	\frac{qb+(1-q)}{\ex[Z]}xp_x, & \quad y\in H_{h(i)} \\
	\frac{1-q}{\ex[Z]}xp_x, & \quad y\in H_j, j\ne h(i).
\end{array}\right.\label{cond-pmf}
\end{align}
From (\ref{marginal-pmf}) and (\ref{cond-pmf}) we obtain
\begin{align}
&\pr(X=x, Y=y)=\pr(X=x | Y=y)\pr(Y=y)\nonumber\\
&=\left\{\begin{array}{ll}
\frac{qb+(1-q)}{(\ex[Z])^2}xy p_xp_y,
& \quad x\in H_i, y\in H_{h(i)}\\
\frac{1-q}{(\ex[Z])^2}xy p_xp_y, &
\quad x\in H_i, y\in H_j,  j\ne h(i)
\end{array}\right. \label{joint-pmf}
\end{align}

We next analyze the expected value of $Y$ and  the product $XY$, respectively.  From (\ref{marginal-pmf}), we obtain
\begin{equation}
\ex[Y]=\sum_{y}y\Pr(Y=y)=\sum_{j=1}^{b}\sum_{y\in H_j}\frac{y^2 p_y}{\ex[Z]}=\frac{1}{\ex[Z]}\sum_{j=1}^{b}u_j.\label{eY0}
\end{equation}
where
\begin{equation}
u_j = \sum_{y\in H_j} y^2 p_y.\label{def-u}
\end{equation}
From (\ref{joint-pmf}), we have 
\begin{align}
\ex[XY]&=\sum_{x}\sum_{y}xy\pr(X=x, Y=y)\nonumber\\
&=\frac{1-q}{(\ex[Z])^2}\sum_{i=1}^{b}\sum_{j=1}^{b}u_{i}u_{j}+\frac{qb}{(\ex[Z])^2}\sum_{i=1}^{b}u_{i}u_{h(i)}.\label{eXY0}
\end{align}
We now consider $\ex[Y | X]$.
Denote the conditional expectation $\ex[Y | X=x]$ by $g(x)$.  Assume that
$x\in H_i$ for some $i$.  From (\ref{cond-pmf}),
we have
\begin{align}
g(x)&= \ex[Y | X=x] \nonumber\\
&=\sum_{i=1}^b\frac{(1-q) u_{i}}{\ex[Z]}+\frac{qb u_{h(i)}}{\ex[Z]}. \label{g(x)}
\end{align}

In addition, the analysis of the clustering coefficient needs the probability that two specific vertices
are connected by a regular edge.  Randomly select two vertices, say vertices $A$ and $B$.  Denote
the degrees of $A$ and $B$ by $X_A$ and $X_B$.  Let the blocks
 of $A$ and $B$ be $Q_A$, $Q_B$, respectively.  We now consider the conditional connection probability
\begin{align}
&p_c(A, B)=\nonumber\\
&\pr(\mbox{vertices $A$ and $B$ are connected}\, |\,X_A=k_A, X_B=k_B,\nonumber\\
&\quad Q_A=i, Q_B=j).\label{ccp}
\end{align}
If  $h(i)\ne j$, vertices $A$ and $B$ can only
be connected through a pair of type 2 stubs.  A type 2 stub of vertex $A$ connects to
a type 2 stub of vertex $B$ with probability $(1-q)k_B/(2m(1-q))$, since $B$ has
on average $(1-q) k_B$ type 2 stubs, there are totally $2m(1-q)$ type 2 stubs in the
network, and the connection is randomly selected.  Since vertex $A$ has $(1-q)k_A$
type 2 stubs on average, it follows that
\begin{equation}\label{ccp-type2}
p_c(A, B)=\dfrac{(1-q)^2 k_A k_B}{2m(1-q)}=\dfrac{(1-q) k_A k_B}{2m}.
\end{equation} 
If $h(i)=j$, vertices $A$ and $B$ can be
connected through a pair of type 1 or type 2 stubs.  In this case,
\begin{equation}\label{ccp-type12}
p_c(A, B)= \dfrac{(1-q)^2 k_A k_B}{2m(1-q) }
+\dfrac{q^2 k_A k_B}{2mq/b}=\dfrac{(1-q+qb) k_A k_B}{2m}.
\end{equation}

\section{Pearson degree correlation Coefficient}\label{s:pdcc}
In the section we analyze the Pearson degree correlation of the GCTC model.  Recall that the GCTC model 
can have multiple communities.
Stubs corresponding to vertices are first divided into communities.
In each community, stubs are divided into blocks to create a non-zero Pearson degree
correlation.  This two layer structure of stubs does not change the mathematical
nature on  how Pearson degree correlation being derived.  However, it does increase
the number of cases and the complexity of notations significantly.  For this reason,
we assume that there is one community in this section. 
In this special case, $c=1$, $r_{i}=1$, $h_i(i) = h(i)$ and $a_{d} = a$.

Pearson degree correlation is defined as the Pearson correlation coefficient of degrees at the two
ends of a randomly selected edge.  The GCTC model has two types of edges, regular 
edges and transitive edges.  In this paper
we analyze the correlation coefficient of degrees at the two ends of a randomly selected regular edge.
Randomly select an edge among regular edges in the network.  Let
$X$ and $Y$ be the number of regular edges that the two ends of the edge have.  Let
$X'$ and $Y'$ be the number of transitive edges that the two ends of the edge have. 
We shall analyze the Pearson degree correlation 
\begin{align}
\rho(X+X',Y+Y')\defeq\dfrac{\cov(X+X',Y+Y')}{\sigma_{X+X'}\sigma_{Y+Y'}}
\label{rho}
\end{align}
where $\cov(X+X',Y+Y')$ is the co-variance of random variables $X+X'$ and $Y+Y'$
and $\sigma_{X+X'}$ is the standard deviation of $X+X'$.

Recall that in the generalized configuration model, the degree 
covariance is defined as $\cov(X,Y)=\ex[XY]-\ex[X]\ex[Y]$
and $\sigma_{X}=\ex[X^2]-(\ex[X])^2$. In the GCTC model, the degree 
co-variance is defined as
\begin{align}
&\cov(X+X',Y+Y')\nonumber \\
 & =\left(\ex[XY]-\ex[X]\ex[Y]\right)+\left(\ex[X'Y]-\ex[X']\ex[Y]\right)\nonumber \\
& \quad+\left(\ex[XY']-\ex[X]\ex[Y']\right)+\left(\ex[X'Y']-\ex[X']\ex[Y']\right).\label{cov}
\end{align}
And the product of standard deviations in the denominator of (\ref{rho}) is equal to
\begin{align}
\sigma_{X+X'}\sigma_{Y+Y'} &=\var{X+X'}^2 \nonumber\\
&=\ex[X^2]-(\ex[X])^2+2(\ex[XX']-\ex[X]\ex[X'])\nonumber\\
&\quad+\ex[(X')^2]-(\ex[X'])^2.\label{var}
\end{align}
Compared with $\cov(X,Y)$, $\cov(X+X',Y+Y')$ has additional expected terms related to $X'$ and $Y'$. 
Before we present the derivations of $\ex[X']$, $\ex[X'Y]$  and $\ex[X'Y']$, we first show the following theorem. 
It is one of the main results in this paper.
Its proof is presented in Appendix B at the end of this paper.

\bthe{thm1}
If $h(i)=i$, then 
\begin{equation}
\cov(X+X', Y+Y')\ge \cov(X, Y)\ge 0.\label{thm1-1}
\end{equation}
\ethe

We next analyze the terms needed to compute $\cov(X+X',Y+Y')$ and $\sigma_{X+X'}$.  We observe that all the quantities of these expected terms 
are in the form of $\ex[X^i Y^j (\ex[Y | X])^k(\ex[X | Y])^l]$ for some integers $i$, $j$, $k$ and $l$. Therefore, we present the
analysis of $\cov(X+X',Y+Y')$ in the following three subsections. In Section \ref{s:ex'}, we show the analysis of $\ex[X']$, $\ex[X'Y]$  and $\ex[X'Y']$.
We next analyze the expected value $\ex[X^i Y^j (\ex[Y | X])^k(\ex[X | Y])^l]$ for integers $i$, $j$, $k$ and $l$ in Section \ref{s:basics}. In Section \ref{s:cov}, we
present the final equations of $\cov(X+X',Y+Y')$. Since the analysis of $\sigma_{X+X'}$ is similar to $\cov(X+X',Y+Y')$, we present the analysis of $\sigma_{X+X'}$
in Appendix-C.

\subsection{Analysis of  $\ex[X']$, $\ex[X'Y]$,  and $\ex[X'Y']$}\label{s:ex'}
To analyze $\ex[X']$, $\ex[X'Y]$,  and $\ex[X'Y']$, we shall first analyze the expected value of $X'$. Recall that $X$ and $Y$ are the number
of regular edges that two vertices at the two ends of a randomly selected regular edge.
Let $A$ and $B$ denote the two vertices.
$X'$ and $Y'$ are the number of transitive edges that $A$ and $B$ have, respectively. 
Number the $X$ regular edges such that the first edge connects to $B$. 
Along the $i$-th regular edge of $A$ to reach 
the other side, where $i=2, 3, \ldots, X$, one finds $Y_i$ regular edges. 
A graphical illustration is shown in \rfig{ill1}.

\bfig{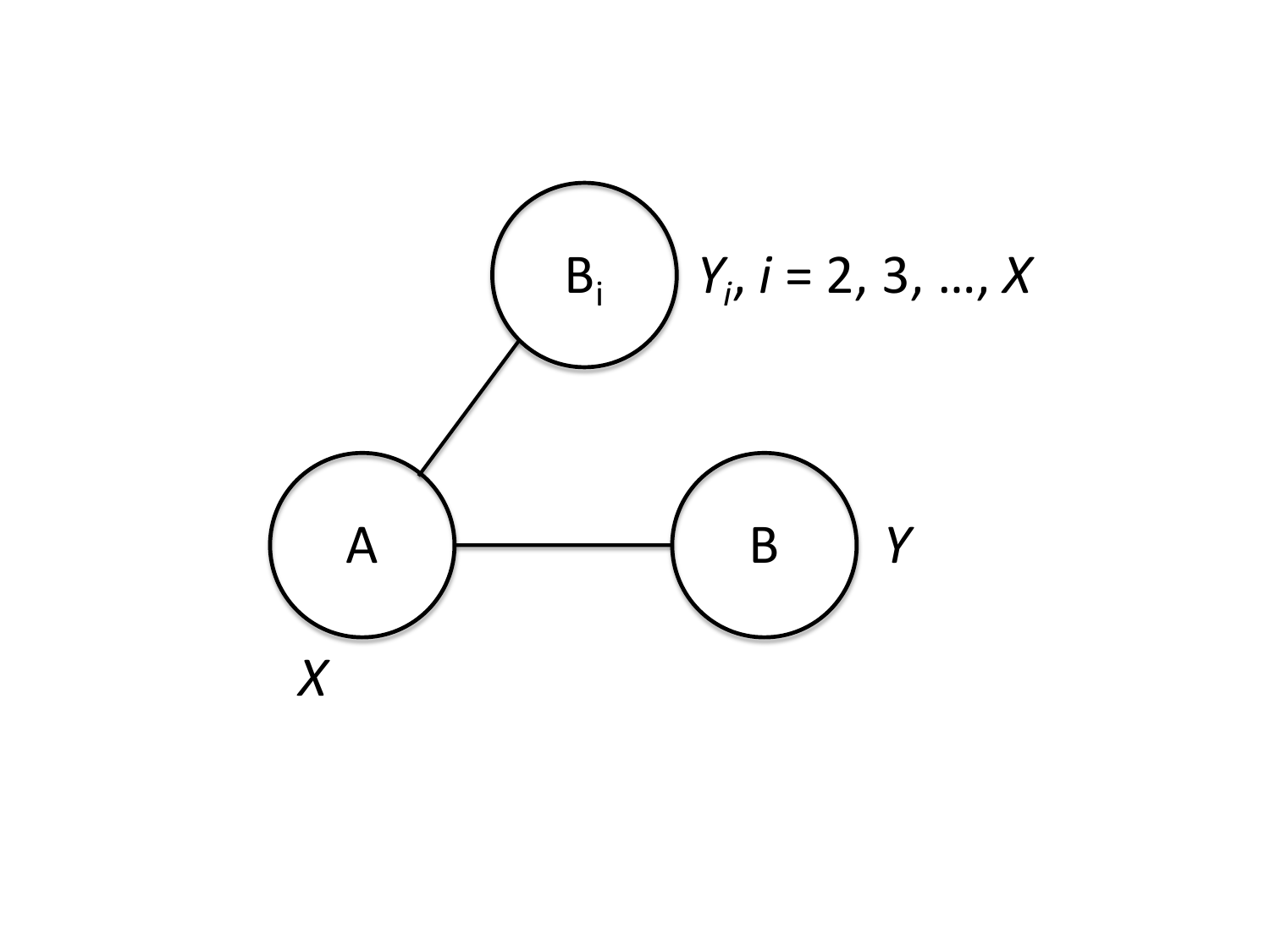}{3.75in}
\efig{ill1}{Edge $AB$ is a randomly selected regular edge.  Vertex $A$ has $X$ regular
	edges. The first edge connects to 
	$B$, which has $Y$ regular edges.  The $i$-th edge connects to vertex $B_i$, which
	has $Y_i$ regular edges.}

Along the first edge, $A$ has $Y-1$ second neighbors and 
along the $i$-th edge, $A$ has $Y_i-1$ second neighbors.  Totally, vertex $A$ has
\begin{equation}
Y-1+\sum_{i=2}^{X} (Y_i-1)\label{nsn}
\end{equation}
second neighbors.  The total number of second neighbors in (\ref{nsn}) can be
overestimated, as some second neighbors can be counted more than once.  However,
as the network size is large, the error is asymptotically small.
We also remark that random variables $Y_i$, $i\le 2$, are identically distributed
for large networks.  Their common distribution is the same as that of
$Y$.
The number of transitive edges that vertex $A$ has, given $X$ and $Y$, is
\begin{equation}\label{nte}
X'=\sum_{j=1}^{Y-1}B_{1,j}+\sum_{i=2}^X \sum_{j=1}^{Y_i-1} B_{ij},
\end{equation}
where $\{B_{ij}: i\ge 1, j\ge 1\}$
is a doubly indexed sequence of independent and identically distributed Bernoulli random
variables with success probability $a$. 
The conditional expectation of $X'$, given $X$ and $Y$, is
\begin{align}
\ex[X'|X,Y]&=\ex\left[\sum_{j=1}^{Y-1}B_{1,j}+\sum_{i=2}^X 
\sum_{j=1}^{Y_i-1} B_{ij}\Biggl|
X,Y\right]\nonumber\\
&=a\cdot(Y-1)+\sum_{i=2}^X a\cdot \ex[Y_i-1 | X].\label{X'-der}
\end{align}
Since $Y_i$ and $Y$ are identically distributed for all $i$, the preceding equation
can be rewritten as
\begin{equation}\label{X'-der2}
\ex[X'|X,Y]= a\left(Y-1+(X-1)\ex[Y|X]-(X-1)\right).
\end{equation}
Taking expectation with respect to $X$ and $Y$, we have
\begin{align}
\ex[X']&=a(\ex[Y]-1+\ex[XY]-\ex[Y]-\ex[X]+1)\nonumber\\
&=a(\ex[XY]-\ex[X]).\label{X'}
\end{align}

Next, we analyze $\ex[X'Y]$. From (\ref{nte}) and similar to (\ref{X'-der}), we have
\begin{align*}
\ex[X'Y|X,Y]&=\ex\left[\left(\sum_{j=1}^{Y-1}B_{1,j}+\sum_{i=2}^X 
\sum_{j=1}^{Y_i-1} B_{ij}\right)Y\Biggl|
X,Y\right]\nonumber\\
&=a((Y-1)Y+  (X-1) Y\cdot \ex[Y-1 | X])\nonumber\\
&=a(Y^2-XY+XY\ex[Y | X] -Y\ex[Y | X]).
\end{align*}
It follows that 
\begin{equation}
\ex[X'Y] = a(\ex[Y^2]-\ex[XY]+\ex[XY\ex[Y | X] ]-\ex[Y\ex[Y | X] ]).\label{X'Y}
\end{equation}
We see that the terms on the right of the preceding expression
are in the form of $\ex[X^i Y^j (\ex[Y | X])^k(\ex[X | Y])^l]$ for some integers $i$, $j$, $k$ and $l$. 
We shall analyze these terms in the next subsection.

Finally, we analyze $\ex[X'Y']$.  From (\ref{nte}), we have
\begin{align*}
\ex[X'Y'|X,Y]&=\ex\left[\left(\sum_{j=1}^{Y-1}B_{1,j}+\sum_{i=2}^X 
\sum_{j=1}^{Y_i-1} B_{ij}\right)\right.\nonumber\\
&\quad\left.\cdot\left(\sum_{j=1}^{X-1}C_{1,j}+\sum_{i=2}^Y 
\sum_{j=1}^{X_i-1} C_{ij}\right)\Biggl|X,Y\right],
\end{align*}
where $\{C_{ij}: i\ge 1, j\ge 1\}$
is a doubly indexed sequence of independent and identically distributed Bernoulli random
variables with success probability $a$.   Doubly indexed sequences $\{B_{ij}\}$ and
$\{C_{ij}\}$ are independent.  Thus, we have
\begin{align*}
\ex[X'Y'|X,Y] &= a^2(Y-1+(X-1)\ex[Y|X]-(X-1)) \\
&\quad \cdot (X-1+(Y-1)\ex[X|Y]-(Y-1)).
\end{align*}
Then, we obtain 
\begin{align}
\ex[X'Y'] &= a^2(-2\ex[X^2]+2\ex[X^2Y]\nonumber \\
&\quad-2\ex[XY\ex[Y|X]]+2\ex[Y\ex[Y|X]]\nonumber \\
&\quad+\ex[\ex[Y|X]\ex[X|Y]]\nonumber \\
&\quad -2\ex[Y\ex[Y|X]\ex[X|Y]]\nonumber \\
&\quad+\ex[XY\ex[Y|X]\ex[X|Y]]).\label{X'Y'}
\end{align}

\subsection{Analysis of $\ex[X^i Y^j (\ex[Y | X])^k(\ex[X | Y])^l]$}\label{s:basics}

We have observed that all quantities of expected values in (\ref{cov}) 
and (\ref{var}) are in the form of
$\ex[X^i Y^j (\ex[Y | X])^k(\ex[X | Y])^l]$ for some integers $i$, $j$, $k$ and $l$. 
Instead of tediously presenting the derivations of all expectation terms needed to compute the covariance and
the variance, we choose a more complex term, that is $\ex[X^2(\ex[Y|X])^2]$, and derive it in this subsection.
The derivation of other terms is similar, and is omitted.  We simply present the result in
Appendix A.

With (\ref{g(x)}), we have
\begin{align*}
\ex[X^2(\ex[Y|X])^2]&= \sum_x(x g(x))^2 \pr(X=x) \\
&=\sum_{i=1}^b \sum_{x\in H_i} \left(\sum_{i=1}^b\frac{(1-q) u_{i}}{\ex[Z]}+\frac{qb u_{h(i)}}{\ex[Z]}\right)^2 \\
&\quad \cdot x^2\cdot\frac{x p_x}{\ex[Z]} \\
&= (1-q)^2\frac{(\ex[Z^2])^2\ex[Z^3]}{(\ex[Z])^3}\\
&\quad+2(1-q)qb\frac{\ex[Z^2]}{(\ex[Z])^3}\sum_{i=1}^bu_{h(i)}t_i\\
&\quad+q^2b^2\frac{1}{(\ex[Z])^3}\sum_{i=1}^b(u_{h(i)})^2t_i,
\end{align*}
where
\begin{align}
t_{i} &=\sum_{x\in H_{i}}x^{3}p_{x}.\label{def-t}
\end{align}

\subsection{Analysis of $\cov(X+X',Y+Y')$}\label{s:cov}
Finally, substituting (\ref{X}), (\ref{X'}), (\ref{XY}), (\ref{X'Y}), and (\ref{X'Y'})
into (\ref{cov}), we obtain
\begin{equation}
\cov(X+X',Y+Y') =\alpha_{0} + \sum_{i=1}^5 \beta_i W_i,\label{cov-final}
\end{equation}
where
\begin{align} 
& \alpha_{0} =2a\frac{a(\ex[Z^{2}]-\ex[Z])+\ex[Z]}{(\ex[Z])^{3}}\left(\ex[Z]\ex[Z^3]-(\ex[Z^{2}])^{2}\right)\nonumber \\
& \beta_{1}= \frac{q}{(\ex[Z])^2}+2a \frac{q}{(\ex[Z])^{2}}\left(\frac{(1-q)\ex[Z^{2}]}{\ex[Z]}-1\right)\nonumber\\
&\ \ +a^{2}\frac{q\left(((1-q)\ex[Z^{2}]+q\ex[Z])^{2}-2(2-q^{2})\ex[Z^{2}]\ex[Z]\right)}{(\ex[Z])^{4}}\nonumber \\
& \beta_{2} =2a^{2}\frac{q}{(\ex[Z])^{2}}\nonumber \\
& \beta_{3} =-2a\frac{q^{2}}{(\ex[Z])^{2}}\left((1-a)+a\frac{(1-q)\ex[Z^2]}{\ex[Z]}\right) 
\nonumber\\
& \beta_{4} =2a\frac{q^{2}b}{(\ex[Z])^{3}}\left((1-a)-aq+a\frac{(1-q)\ex[Z^2]}{\ex[Z]}
\right) \nonumber\\
& \beta_{5} = a^{2}\frac{q^{3}b^{2}}{(\ex[Z])^{4}} \nonumber \\
&W_1=b\sum_{i=1}^b u_i u_{h(i)}-\sum_{i=1}^bu_i\sum_{j=1}^b  u_j\label{W_1} \\
&W_2=b\sum_{i=1}^b t_i u_{h(i)}-\sum_{i=1}^bu_i \sum_{j=1}^b t_j\label{W_2}\\
&W_3= b\sum_{i=1}^{b}u_{i}u_{i}-
\sum_{i=1}^{b}u_{i}\sum_{j=1}^{b}u_{j}\label{W_3}\\
&W_4= b\sum_{i=1}^{b}u_iu_{h(i)}u_{i}-
\sum_{i=1}^{b}u_{i}u_{h(i)}\sum_{j=1}^{b}u_{j}\label{W_4}\\
&W_5=b\sum_{i=1}^{b}u_{i}u_{h(i)}u_{i}u_{h(i)}-\sum_{i=1}^{b}u_{i}u_{h(i)}\sum_{j=1}^{b}u_{j}u_{h(j)}.\label{W_5}
\end{align} 
In (\ref{W_1}) and (\ref{W_2}), sequences $\{u_i\}$ and $\{t_i\}$ are defined in (\ref{def-u})  
and (\ref{def-t}), respectively.

\section{Clustering coefficient}\label{s:lcc}

In this section we analyze the local clustering coefficient of
the GCTC model.  We remark that in this section we study the clustering coefficient
of a special case in which there is only one community.  We also remark that
it is easy to extend the analysis to GCTC models with more than one community.
We choose to present the result of the special case in order to keep notational simplicity.

Let $A$ be a randomly
selected vertex in a random network.  Let $k$ be the degree
of $A$.  The local clustering coefficient of vertex $A$ is defined as
\begin{align}
&C_A(k)=\frac{\mbox{number of connected pairs of neighbors of $A$}}
{\mbox{number of pairs of neighbors of $A$}}\nonumber\\
&=\frac{\mbox{number of connected pairs of neighbors of $A$}}
{k(k-1)/2}.\label{config-lcc-def}
\end{align}
If $k\le 1$, $C_A(k)$ is defined to be zero. We distinguish between regular edges and transitive edges.   Assume
that vertex $A$ has $k$ regular edges and $k^\prime$ transitive edges.  
We denote the local clustering coefficient of $A$ by $C_A(k, k')$.
Let the $k$ vertices connected with
$A$ by regular edges be denoted by $U_1, U_2, \ldots, U_k$.  Let
the $k^\prime$ vertices connected with $A$ by transitive edges be
denoted by $V_1, V_2, \ldots, V_{k^\prime}$.

We analyze the numerator of (\ref{config-lcc-def}).  Obviously, if $k=0$,
$C_A(0, k^\prime)=0$.  For general $k\ge 1$ and $k^\prime\ge 0$, we claim that
\begin{equation}\label{cc-result}
C_A(k,k')=\frac{\displaystyle\binomcoeff{k}{2}a+k^\prime+\binomcoeff{k^\prime}{2}\times \frac{a}{k}}
{\binomcoeff{k+k^\prime}{2}},
\end{equation}
with the convention that
\[
\binomcoeff{i}{j}=0\ \mbox{if $i<j$}.
\]

To analyze (\ref{cc-result}) we consider six types of triangles as shown in
\rfig{lcc}. We consider type 1 triangles shown in panel (a) of
\rfig{lcc}.  The expected number of type 1 triangles is
\[
\sum_{k_1, k_2} \left(\begin{array}{c} k \\ 2 \end{array}
\right) p_c(U_1, U_2) p_{k_1} p_{k_2},
\]
where $k_1$ and $k_2$ are the degrees of vertices $U_1$ and $U_2$,
respectively.  
Since $2m=n \ex[Z]$, it follows that the expected number of
type 1 triangles in the last expression approaches to zero as
the network size $n$ is large.  Note that
type 5 triangles in panel (e)
of \rfig{lcc} also require vertices $U_1$ and $U_2$ be
connected by regular edges.  By the same argument, it is easy to
see that the expected number of type 5 triangles also
goes to zero as the network gets large.

Now we consider the second type of triangles shown in panel (b)
of \rfig{lcc}.  Vertices $U_1$ and $U_2$ are connected by a
transitive edge.  This transitive edge is formed because vertice
$U_2$ is an unconnected second neighbor of $U_1$ through vertex $A$.
Thus, the expected number of type 2 triangles is
\[
\sum_{k_1, k_2} \left(\begin{array}{c} k \\ 2 \end{array}
\right)\left(1- p_c(U_1, U_2) p_{k_1} p_{k_2}\right)\cdot a
=\left(\begin{array}{c} k \\ 2 \end{array}\right)\cdot a.
\]
This is the first term in the numerator of (\ref{cc-result}).

We next analyze type 3 triangles shown in panel (c) of \rfig{lcc}. Note that transitive edge $AV_1$  can be formed in two types of event. The first type of event is  the successful event of random triadic closure of the connected triples of $A$, $V_1$ and  $U_1$.  The second type of event is the successful event of random triadic closure of the connected triples of $A$, $V_1$ and  one of the first neighbors of A  in the
set $\{U_1, U_2, \ldots, U_k\}$ except $U_1$. If the transitive edge $AV_1$  in type 3 triangle is formed from the first type of event, the number of type 3 is $k'$. If the  transitive edge $AV_1$  is formed from the second type of event, to form a type 3 triangle, $U_1$ and  $V_1$ need to be connected by a regular edge. From the analysis of type 1 triangles, we know that the connected probability of $U_1$ and  $V_1$ approaches to zero as the network size $n$ is large. To sum, the expected number of type 3 triangles is  $k'$, which is the second term in  the numerator of (\ref{cc-result}).

Now we consider type 4 triangles shown in panel (d)
of \rfig{lcc}.  In order to form a transitive edge between
$U_1$ and $V_1$, these two vertices must have at least one common neighbor
by regular edges.  Besides vertices $A$, $U_1$ and $V_1$, there
are $n-3$ vertices in the network.  Let $E_{n-3}$ be the event 
that there is at least one vertex in $n-3$ vertices that connects
to both $U_1$ and $V_1$.  Then, the expected number of type 4
triangles is
\begin{equation}
k k' a \pr(E_{n-3})=k k' a (1-\pr(E_{n-3}^c)),\label{ntt4}
\end{equation}
where $E_{n-3}^c$ is the complement of event $E_{n-3}$. Denote the
$n-3$ vertices by vertices $1, 2, \ldots, n-3$.
\begin{align}
\pr(E_{n-3}^c)=&\sum_{u, v, k_1, k_2, \ldots, k_{n-3}}
\prod_{j=1}^{n-3} (1-p_c(j,U_1))(1-p_c(j, V_1))\nonumber\\
&\times p_u p_v p_{k_1} p_{k_2}\cdots p_{k_{n-3}},\label{pEc}
\end{align}
where $u$ and $v$ are the degrees of $U_1$ and $V_1$, and
$k_j$ is the degree of vertex $j$ for $j=1, 2, \ldots, n-3$.  
To evaluate $\pr(E_{n-3})$,
we substitute (\ref{ccp-type2}) or
(\ref{ccp-type12}) into (\ref{pEc}).  For example, denote $b_{U_1}$, $b_{V_1}$ and $b_{j}$ as the block index of 
 $U_1$, $V_1$ and $j$, respectively. Suppose $b_{U_1} \ne h(b_{j})$ and $b_{V_1} \ne h(b_{j})$, we substitute (\ref{ccp-type2}) into (\ref{pEc}) and have 
\begin{align}
&\pr(E_{n-3})=1-\nonumber\\
&\sum_{u, v, k_1, k_2, \ldots, k_{n-3}}
\prod_{j=1}^{n-3} \prod_{j=1}^{n-3} \left(1-\frac{(1-q)u k_j}{2m}
\cdot\frac{(1-q)v(k_j-1)}{2m}\right) \nonumber\\
&\times p_u p_v p_{k_1}p_{k_2}\cdots p_{k_{n-3}}\nonumber\\
&=1-\sum_{u,v}\left(1-\frac{(1-q)^2 u v (\ex[Z^2]-\ex[Z])}{(n \ex[Z])^2}
\right)^{n-3}p_u p_v \nonumber\\
&\le 1-\left(1-\frac{(1-q)^2 (\ex[Z])^2(\ex[Z^2]-\ex[Z])}{(n\ex[Z])^2}
\right)^{n-3}\label{fe} \\
&\to 0\ \mbox{as $n\to\infty$,}\nonumber
\end{align}
where inequality (\ref{fe}) is due to Jensen's inequality \cite{FelW.B}.
It follows that the expected number of type 4 triangles is zero 
in large networks.  Note that substitution of (\ref{ccp-type12}) into 
(\ref{pEc}) leads to the same result, {\em
i.e.} the expected number of type 4 triangles is zero in large networks.

Finally, we consider the expected number of type 6 triangles shown
in panel (f) of \rfig{lcc}.  Note that to form transitive edges
$AV_1$ and $AV_2$, vertices $V_1$ and $V_2$ must be unconnected
second neighbors of $A$ through some first neighbors of $A$ in the
set $\{U_1, U_2, \ldots, U_k\}$.  There are two cases. 
In the first case shown in panel (a) of \rfig{t6t}, $V_1$ and $V_2$ have
distinct common neighbors with $A$.  
Vertices $V_1$ and $V_2$ randomly and independently select first
neighbors from the set $\{U_1, U_2, \ldots, U_k\}$.  The probability
that their selections are distinct is
\[
(k-1)/k.
\]
Thus, the expected number of type 6 triangles in the first case
is 
\[
\left(\begin{array}{c}k' \\ 2 \end{array}\right)
\cdot\dfrac{k-1}{k}\cdot a\cdot \pr(E_{n-5}).
\]
By the same argument in (\ref{fe}), it is easy to show that
the quantity above goes to zero as $n$ goes to infinity.
Now we consider the second case shown panel (b) of \rfig{t6t}. Vertices $V_1$ and $V_2$ share
a common first neighbor $U_i$ with $A$.  The probability that
the random selections of $V_1$ and $V_2$ are the same is $1/k$.
Thus, the expected number of type 6 triangles is
\[
\left(\begin{array}{c}k' \\ 2 \end{array}\right)
\cdot\dfrac{a}{k}.
\]
This is the third term in the numerator on the right side of (\ref{cc-result}).

\bfig{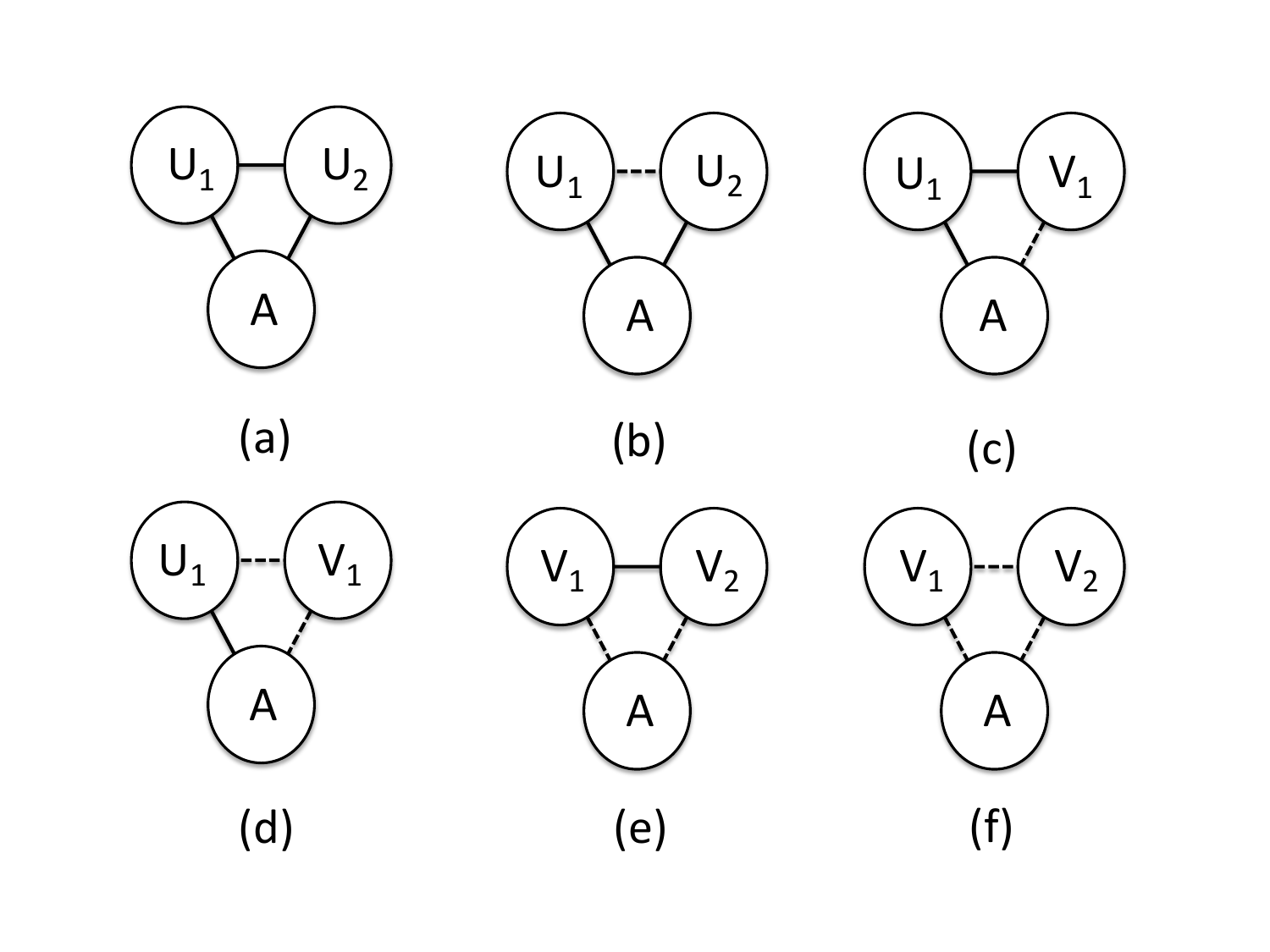}{3.75in}
\efig{lcc}{Six types of triangles.  Solid lines denote regular edges
of the GCTC model.  Dashed lines denote 
transitive edges due to triadic closure operations.}

\bfig{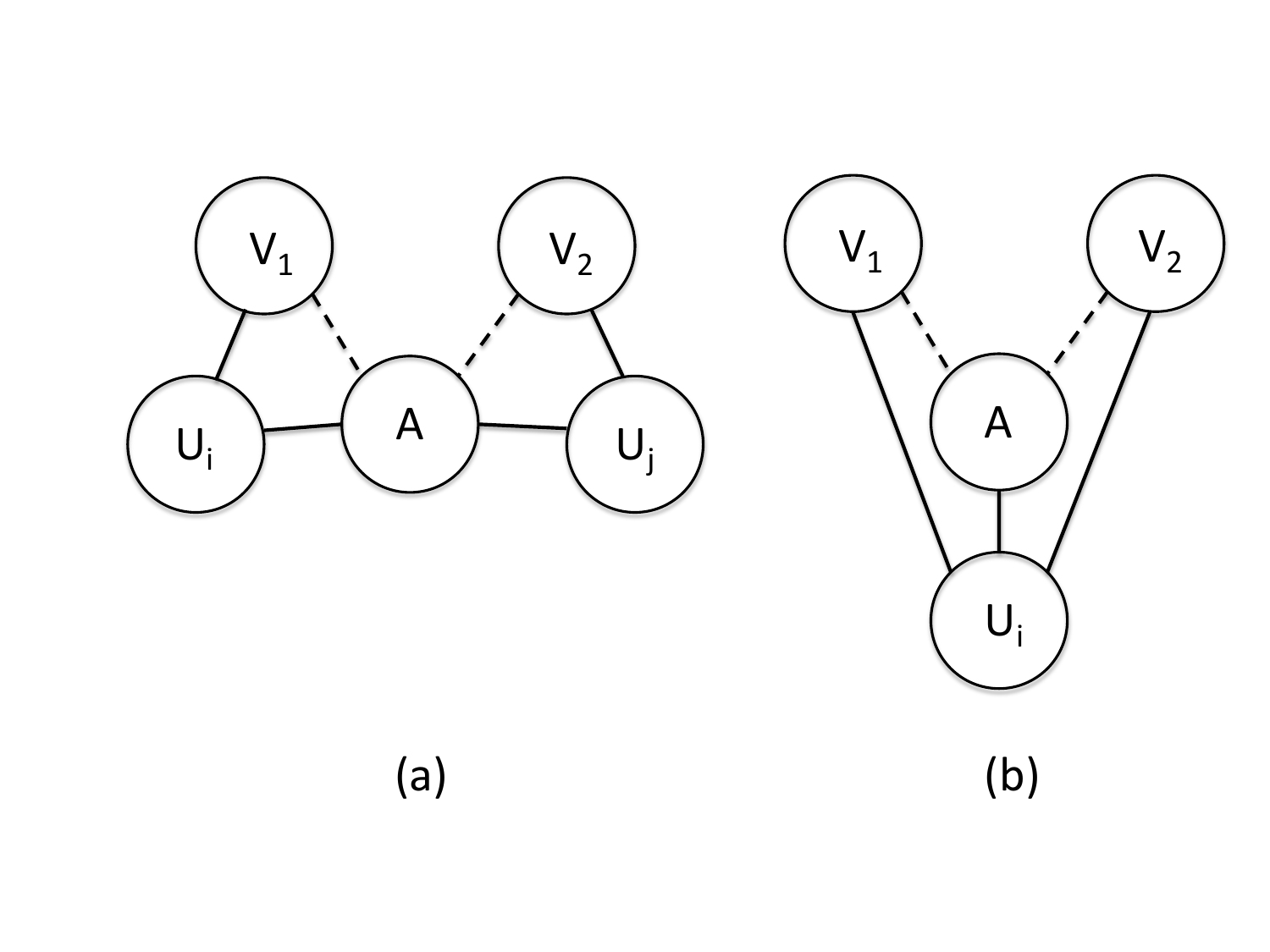}{3.75in}
\efig{t6t}{Type 6 triangles. Vertices $V_1$ and $V_2$ must be unconnected second
neighbors of $A$ through some first neighbors $U_i$ and $U_j$.  In panel
(a), the two first neighbors of $A$ are distinct.  In panel (b), vertices
$V_1$ and $V_2$ have a common first neighbor of $A$.}

\section{Simulation Results}\label{s:numerical}

	In this section we present our numerical and simulation results.
We present our results in four subsections.  
First, since our closed form expressions for the Pearson degree 
correlation coefficient and the clustering coefficient are quite
complicated, we verify their correctness by comparing numerical
results with simulation results in Section \ref{sim:sim}. 
In Section \ref{s: syn}
we model four real-world networks by GCTC networks.  To make a comparison we
also model the same real-world networks using SBM, LFR and ABCD models.
In Section \ref{s:app}, we study whether the GCTC model is suitable
to be a benchmark model for community detection algorithms.
Finally, in Section \ref{s:app2} we simulate influence diffusion
in GCTC networks and compare the result with that in a real-world network.

\subsection{Correctness of Eqs. (\ref{cov-final}) and (\ref{cc-result})}\label{sim:sim}
Since our closed form expressions in Eqs. (\ref{cov-final}) and 
(\ref{cc-result}) are quite complicated, we verify their correctness 
by comparing their numerical results with simulation results.  
Recall that we assume $c=1$, {\em i.e.} there is only one community in the
derivation of Pearson degree correlation coefficient and clustering 
coefficient in Sections \ref{s:pdcc} and \ref{s:lcc}.	We drop subscripts
and use notations $b$ and $q$ to denote the number of blocks and the
fraction of type 1 stubs.  In our experiment, we assume that there
are 10000 vertices.  We sample a power law degree distribution to generate
one degree sequence. For this degree sequence, we randomly construct fifty
GCTC networks.  We calculate the Pearson degree correlation coefficients
and the clustering coefficients of the fifty networks and take an average.
We choose $b=2$.
	
We first consider positive degree correlation and assume that blocks are 
associated with each other by permutation $h(i)=i$.  We present
the numerical calculation and simulation of covariance $\cov(X+X', Y+Y')$
as a function of $q$
in \rfig{covq_assor}. Then, we examine Pearson degree correlation
coefficient $\rho(X+X', Y+Y')$ for permutation $h(i)=i$ and
permutation $h(i)=b+1-i$.  The result is shown in \rfig{cor}.
Note that without triadic closure operations permutation $h(i)=b+1-i$ would
generate disassortatively mixed networks \cite{Lee2019}.  \rfig{cor}
shows that with triadic closure operations and small value of $q$, permutation 
$h(i)=b+1-i$ could generate assortatively mixed graphs.  
Finally, we notice that the numerical results agree very well with the
simulation results in \rfig{covq_assor} and \rfig{cor}.

\bfig{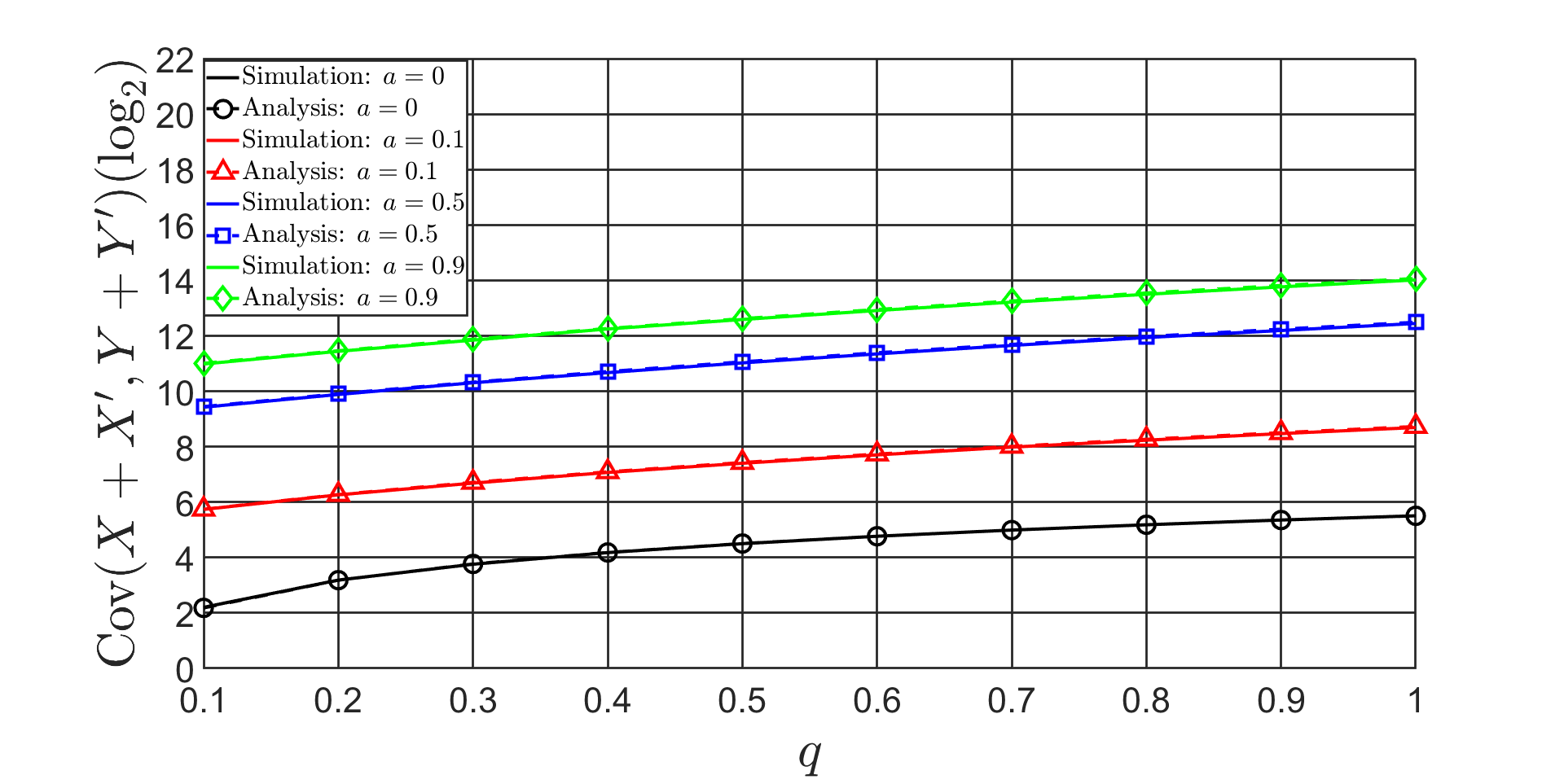}{3.8in}
\efig{covq_assor}{Plot of  co-variance versus $q$ with $h(i)=i$. }		
\bfig{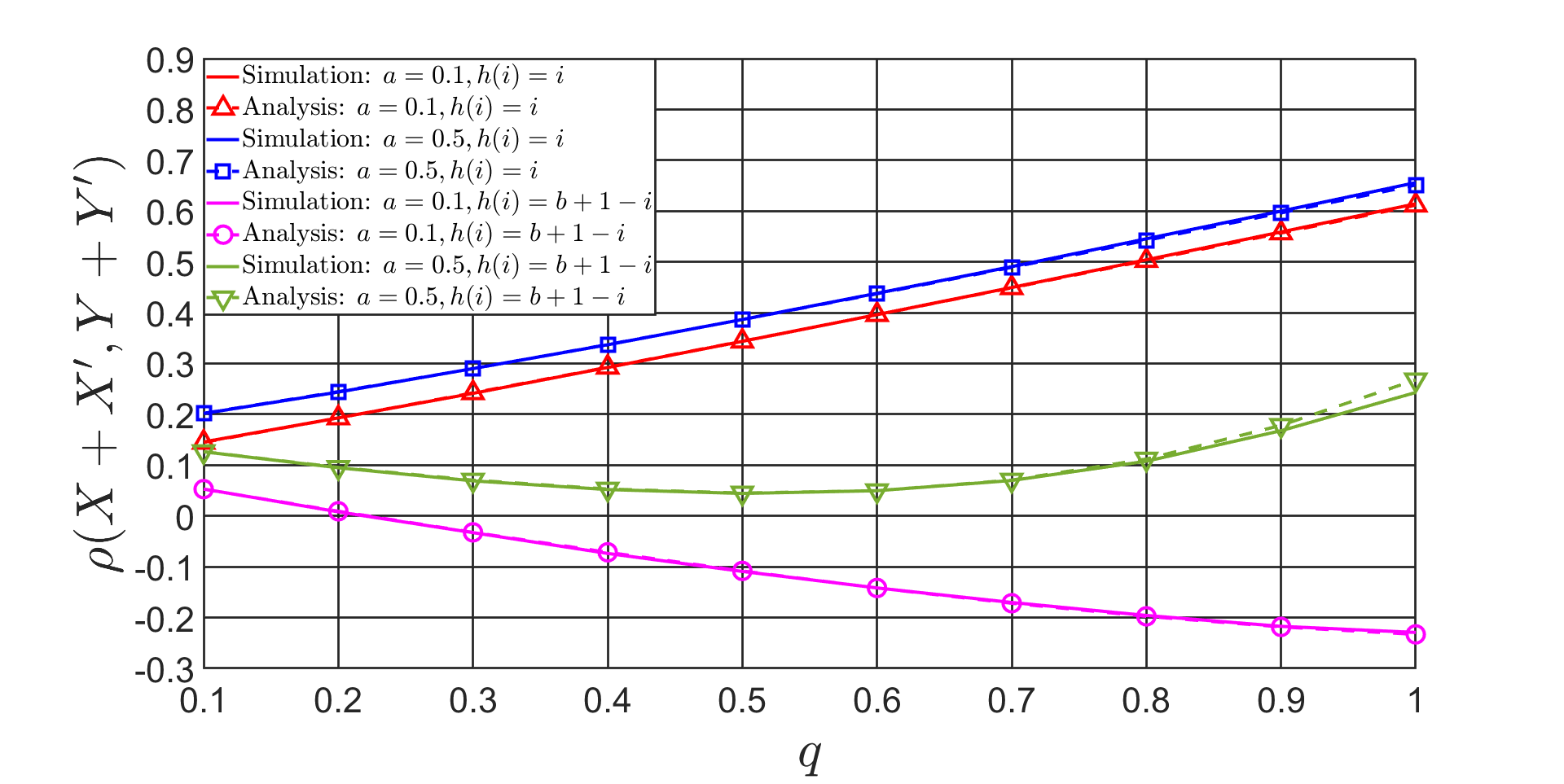}{3.8in}
\efig{cor}{Plot of Pearson degree correlation versus $q$.}

Next, we numerically compute and simulate the clustering coefficient of the 
GCTC graph. The clustering coefficient is calculated by taking an average 
of the clustering coefficients of vertices in the network.
The results are shown in \rfig{sim_lcc}. We find that the simulation result and the
numerical calculation of Eq. (\ref{cc-result}) are very close. 
We notice from \rfig{sim_lcc} that the clustering coefficient is increasing with
$a$, which implies that triadic closure operations increase the transitivity
of the network. We also notice that the range of clustering coefficient is somewhat
narrow for a relatively wide range of $a$ and $q$.


\bfig{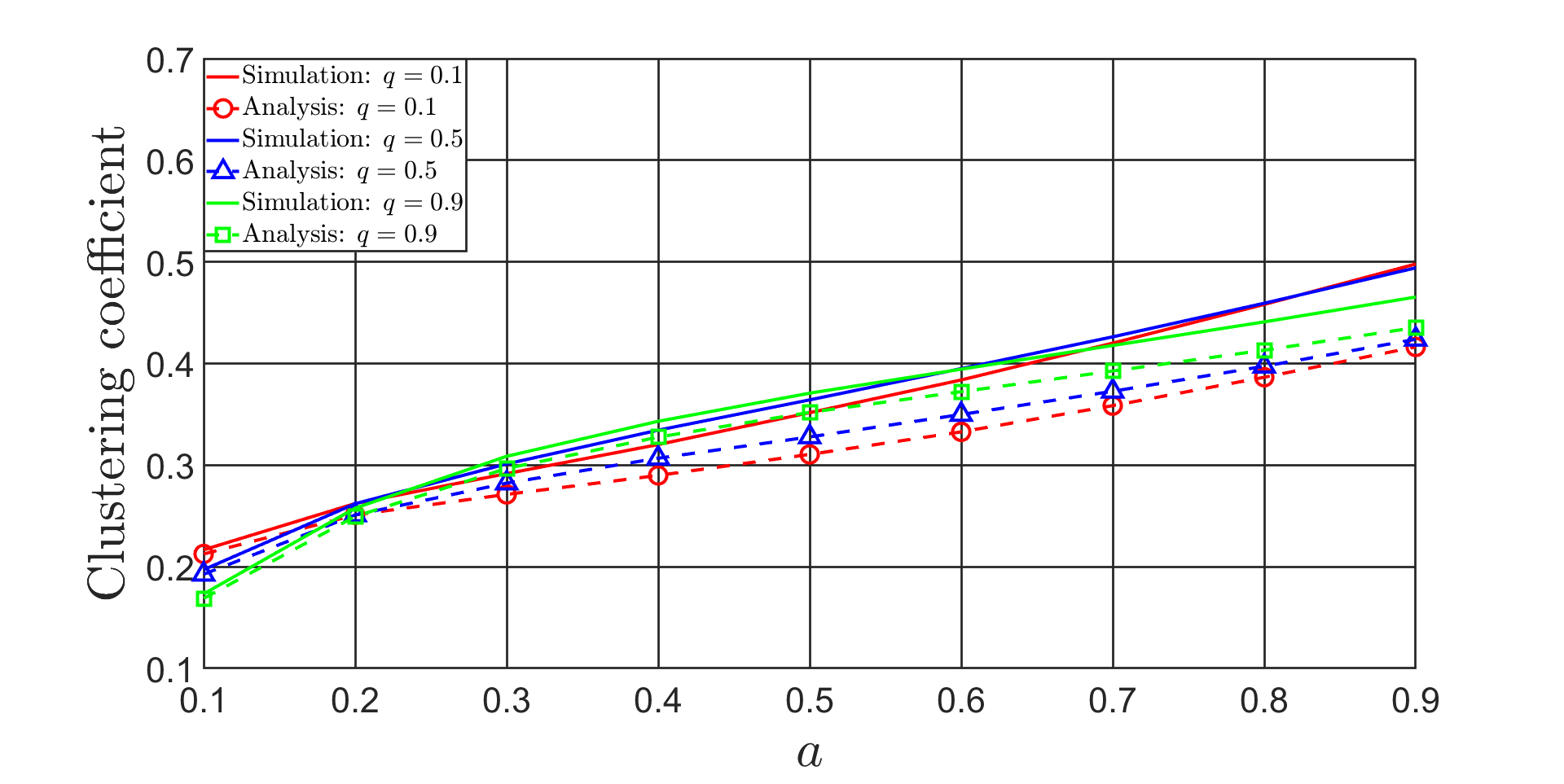}{3.8in}
\efig{sim_lcc}{Plot of clustering coefficient versus $a$ with $h(i)=i$.}

\subsection {Modeling real-world networks}\label{s: syn}

In this section we would like to study if it is possible to choose a proper set
of parameters such that the degree distribution, Pearson degree correlation coefficient
and the clustering coefficient of the GCTC model match reasonably well with those
of a real-life network.  To study this problem, we choose four networks collected
in the real life.  Since we will study community detection algorithms on real-world networks
in the next section,
we choose real-world networks with a known community structure.  Specifically,
the four real-world networks that we choose are the Amazon network, the Email network, 
the Ogbn-arxiv network and the DBLP network.  The Ogbn-arxiv data 
can be obtained from the OGB 
website\footnote{\href{https://ogb.stanford.edu/docs/nodeprop}{https://ogb.stanford.edu/docs/nodeprop}}.
The other three network data are available at the 
Stanford website\footnote{\href{https://snap.stanford.edu/data/}{https://snap.stanford.edu/data/}}.
These four datasets need pre-processing before they can be used.  The Amazon network
has multiple overlapped communities.  We keep only one community and delete all other vertices.
As a result, it has only one community.  For the Email network, we keep communities
with more than fifty vertices, and delete smaller communities.  For the Ogbn-arxiv 
network, we only keep two largest communities whose sizes are smaller than
ten thousand vertices.  For the DBLP network, we keep the largest three communities.

We construct GCTC graphs with Algorithm \ref{alg1} using properly selected
parameters to match the performance of the four real-world graphs.  The Pearson
degree correlation coefficient, the clustering coefficient, the average length
of shortest paths between randomly selected vertices, among others, are shown
in Table \ref{tbl1}.  Note that due to triadic closure operations, the total
number of edges $m$ and the expected degree $\ex[Z]$ are not identically equal
to those of the real-world networks.  From column 7 and column 8, we see
that the Pearson degree correlation coefficient and the clustering coefficient
of the GCTC model match quite well with those of the four real-life networks.

To make a comparison, we also simulate three prevalent benchmark models
that are often used to evaluate community detection algorithms.  They are 
the SBM, the LFR model and the ABCD model.  We briefly review these three models.
The simplest SBM is a multi-graph with a given number of
communities. Each vertex is assigned to a community.  Undirected edges are 
placed independently between vertex pairs with probabilities that are 
only a function of the community
membership of the vertices \cite{Holland1983}.  As a result, each vertex in an
SBM has a Poisson distribution for its degree.  To fit an SBM with an empirical
network collected in the real world, one typically formulates a maximum likelihood
problem to determine the parameters in the Poisson distributions.  
Unfortunately, the traditional SBM does not work well in the sense that
it can not fit well with a wide range of network data. 
Karrer \etal \cite{Karrer2011} proposed a degree-corrected version of the SBMs.
In the degree-corrected version, each vertex is associated with a new parameter, with
which the parameters of Poisson distributions are multiplied and thus corrected.  A maximum
likelihood problem solves both the new parameters as well as the parameters of
the Poisson distributions.  We refer the reader to \cite{Karrer2011} for more
details.  In this paper, we choose the degree-corrected SBMs to fit with the four
real-world networks. Next, we briefly review LFR model.
An LFR graph is constructed first by sampling
a power law distribution for a sequence of $n$ degrees.  Then a sequence of community 
sizes is sampled from another power law distribution.  Clearly, the sum of all
community sizes must be equal to the total number of vertices.  
Randomly assign each vertex to one community. 
Users of an LFR model also need to determine another parameter, the fraction of
edges that connect two vertices in two distinct communities.  
Let $\mu$ denote this parameter.
Each vertex is randomly connected by a fraction $1-\mu$ of its links with vertices 
within its community and a fraction $\mu$ of its links with the other vertices of the network.
Finally, we review ABCD model.
The acronym ABCD stands for Artificial Benchmark for Community Detection.
It is a modification of the LFR model in an attempt to make it be constructed faster.
An ABCD model also has a parameter $\xi$ called a mixing parameter.
An ABCD model connects a fraction $1-\xi$ of its edges to other vertices in
the same community, and connects a fraction of $\xi$ edges to vertices globally
including the vertices in the same community.  Finally, we mention that the construction
algorithm for LFR networks has a complexity of $O(m^2)$.
The construction algorithm for GCTC networks has a complexity of $O(n^3)$.

To model a given real-world network, we measure the fraction of edges connecting
two vertices in two distinct communities.  We call this quantity the mixing
parameter of the real-world network.    
For simplicity, we set $\mu$ to this quantity and use it to construct LFR networks. 
We also use this quantity to calculate the value of $\xi$ according to \cite{Kami_2021}. 
We then use $\xi$ to construct ABCD networks. 
In Table \ref{tbl1}, we present the values of the mixing parameter 
for the real-life networks and simulated graphs.
Moreover, from the entries in columns 7 and 8 in Table \ref{tbl1}, 
we see that the Pearson degree
correlation coefficient and the clustering coefficient of the GCTC model match 
consistently better with those of the real-world networks than the SBM, the LFR model,
and the ABCD model.

\begin{table*}[h!]
	\caption{Network properties of four real-world networks and four random graphs}
	\centering 
\begin{threeparttable}[ht]
	\begin{tabular}{c rrrrrrrrrrrrr} 
		\hline\hline 
		\textbf{Graph} &$n$ &$m$ &$\ex[Z]$ &$c$ & mixing &$\rho$ &$C$
		&$\ell$ &\textbf{Walktrap} &\textbf{Eigen} &\textbf{Fast}&\textbf{IC}&\textbf{LT}\\ [0.1ex] 
		& & & & & parameter & & & & & & & & \\
		\hline 
		\textbf{Amazon} &310&895&5.77  &1&0 &-0.1239  &0.4739&5.77 &-&-&-&0.1867&0.2516 \\[0.5ex]
		GCTC 
		&310 &861&5.55 &1&0&$\textbf{-0.1122}$ &\textbf{0.5076}&\textbf{5.55}&-&-&-&\textbf{0.1323}&\textbf{0.2237}  \\[0.5ex]
		SBM
		&310 &891&5.75 &1&0 &-0.0173 & 0.0300&3.33 &-&-&-&0.2800&0.1581 \\[0.5ex]
		LFR 
		&310 &895&5.77 &1&0 &-0.0169 & 0.0249&3.40 &-&-&-&0.2634&0.1528 \\[0.5ex]
		ABCD 
		&310 &895&5.77 &1&0 &-0.0169 & 0.0249&3.40 &-&-&-&0.2634&0.1528\\[0.5ex]
		\hline 
		\textbf{Email}&225 &1507&13.40&4&0.0902&-0.1025 &0.4787  &2.81&0.9300&0.8432&0.9166&0.6757 &0.3467  \\[0.5ex]
		GCTC &225 &1535&13.64 &4 &0.0803 &$\textbf{-0.0965}$ &\textbf{0.3895}&\textbf{2.77}&\textbf{0.9421}&0.8943&\textbf{0.9319}&0.6991&0.3500  \\[0.5ex]
		SBM &225 &1442&12.82&4 &0.0947&-0.0428 &0.3500&2.72&0.8883&\textbf{0.8353}&0.8840&0.6525&\textbf{0.3456}  \\[0.5ex]
		LFR &225 &1507&13.40&4 &0.1058&-0.1436 &0.3789&2.64&0.9489&0.8757&0.9451&0.6970&0.3510   \\[0.5ex]
		ABCD &225 &1507&13.40&4&0.1056 & -0.1445 &0.3797&2.64&0.9490&0.8724& 0.9446& \textbf{0.6969}&0.3504  \\[0.5ex]
		\hline 
		\textbf{Ogbn-arxiv}  &11315 &32694&5.78&2 &0.0138&0.0922  &0.2508&7.50&0.2326&0.3053&0.3255&0.2191&0.4517\\[0.5ex]
		GCTC 
		&11315 &33256&5.88&2 &0.0159&$\textbf{0.1019}$ &\textbf{0.2110} &\textbf{5.87}&\textbf{0.2534}&\textbf{0.3552}&\textbf{0.3569}&\textbf{0.2548}&\textbf{0.4343}\\[0.5ex]
		SBM &11315 &32693&5.78&2&0.0138&$0.0131$ &0.0041&4.92&0.4636&0.3855&0.3798&0.3021&0.0506  \\[0.5ex]
		LFR &11315  &32726&5.78&2 & 0.0143 &-0.0072 &0.0040&5.08 &0.4435&0.7323&0.4821&0.3005&0.0795\\[0.5ex]
		ABCD &11315  &32728&5.78 &2 & 0.0164 &-0.0035 &0.0041&5.05 & 0.4306& 0.7605&0.4715&0.3006&0.0793\\[0.5ex]
		\hline 
		\textbf{DBLP} &15957 &42943&5.38 &3&0.4780&0.2038  &0.6278&7.75 &0.0674&0.0018&0.0201&0.1199&0.0315 \\[0.5ex]
		GCTC 
		&15957  &42374&5.31 &3&0.4659 &$\textbf{0.2241}$ &\textbf{0.5232} &\textbf{9.56} &\textbf{0.0681}&0.0097&\textbf{0.0158}&\textbf{0.1205}&\textbf{0.0485} \\[0.5ex]
		SBM&15957 &42934&5.38  &3 &0.4782& 0.0113 &0.0012&4.92&0.1231&0.0489&0.0415&0.2782&0.0538 \\[0.5ex]
		LFR&15957 &42963&5.39  &3 &0.4783& -0.0014 &0.0011&5.12&0.0963&0.0019&0.0016&0.2482&0.0647 \\[0.5ex]
		ABCD &15957 &42963&5.39 &3& 0.4848 & -0.0011 &0.0011 & 5.12& 0.0953&\textbf{0.0018}&0.0016&0.2482&0.0656\\[0.5ex]
		\hline 
	\end{tabular}
\begin{tablenotes}
      \small
      \item In this table, $n$ is the total number of vertices and $m$ 
      is the number of edges. $\ex[Z]=2m/n$ is the average degree. 
The number of communities is $c$.  The mixing parameter is the ratio of the number of 
edges cross communities to total number of the edges. Next, $\rho$ is the Pearson degree 
correlation coefficient and $C$ is the average of the local clustering coefficients of 
all vertices in the network.
The average length of the shortest paths between two randomly selected vertices is $\ell$.
The entries in columns \textbf{Walktrap},
\textbf{Eigen}, and \textbf{Fast}
are the NMI values of the walktrap, leading eigenvector and the fast greedy 
algorithms, respectively. 
The entries in columns \textbf{IC} and \textbf{LT} are the influence spread
of the influence diffusion processes.  We choose $p=0.15$ as the probability
of influence in the IC model.  We choose $t=0.35$ as the threshold in the
LT model. In the IC model, we simulate the diffusion 1000 times on each graph 
and take an average.  In both
cascade models, we randomly select ten vertices as seeds in the largest 
communities in the Amazon network and in the Email network.
We randomly select 100 and 200 vertices as seeds in the Ogbn-arxiv network 
and in the DBLP network, respectively.
For the entries in columns $\rho$, $C$, $\ell$, {\bf Walktrap}, {\bf Eigen}, 
{\bf Fast}, {\bf IC} and {\bf LT}, we show in boldface those entries 
that are closest to the corresponding entries for real-world networks.
    \end{tablenotes}
 \end{threeparttable}
\label{tbl1} 
\end{table*} \noindent

\subsection {Community detection} \label{s:app}

One of the possible applications of the GCTC model is to serve as a benchmark model
for community detection algorithms.  In this section we compare the performance
of the GCTC model with three well known benchmark models.  

We test the performance of the GCTC model as a benchmark model using three well known
community detection algorithms.  They are
the walktrap algorithm \cite{Pons2005}, the leading eigenvector algorithm
\cite{Newman2006} and the fast greedy algorithm \cite{clauset2004}.  
We use normalized mutual information (NMI) to measure the performance of a benchmark model.
We briefly state the definition of NMI here and refer the reader to \cite{Karrer2011, Danon2005}
for more details.  Let $n_{ij}$ be the number of vertices in community $i$ in the inferred
community detection and in community $j$ in the ground truth.  Define joint probability
$\Pr(C_1=i, C_2=j)=n_{ij}/n$ that a randomly selected vertex is in $i$ in the inferred
detection and $j$ in the ground truth.  Using this joint probability over the random
variables $C_1$ and $C_2$, the NMI is defined as
\[
NMI(C_1, C_2)=\frac{2MI(C_1, C_2)}{H(C_1)+H(C_2)},
\]
where $MI(C_1, C_2)$ is the mutual information between $C_1$ and $C_2$, $H(C_1)$ is the entropy of random variable
$C_1$, and $H(C_2)$ is the entropy of random variable
$C_2$ .  Higher values of NMI indicate a higher degree of
consistency between the detected structure of communities and the ground truth.
Since the Amazon network has only one community,
we have not applied community detection algorithms on it.  
For each benchmark model, we simulate and generate one thousand graphs.  For each graph, we
apply the three community detection algorithms and compute the NMI values.  We present
the average of the NMI values in columns 10, 11 and 12 in Table \ref{tbl1}.
From our simulation results, we see that the GCTC model performs 
better than the SBM, the LFR model,
and the ABCD benchmark model when they model the Ogbn-arxiv network.  
For the other two real-world networks with leading eigenvector algorithm, the performance
of the GCTC model is not the best.  In fact, the performance of the GCTC model 
was the worst among the four benchmark models when the four models synthesize the
Email network to evaluate the leading eigenvector algorithm.
Since the NMI values of the GCTC model generally agree with those of the
real-world networks and the other three random network models, we conclude
that the GCTC model can serve well as a benchmark model for the evaluation of
community detection algorithms.
In addition, since the GCTC model performs the best in matching its $\rho$ and $C$ with those
of the real-world networks, this study seems to imply that community detection problem
is less sensitive to degree correlation and transitivity of a network.
We observe that the mixing parameter for the DBLP network is very close to 1/2.  
This fact might make the DBLP network very difficult for all community detection algorithms.
Indeed, the NMI values of all the three community detection algorithms are very small.

\subsection {Influence diffusion} \label{s:app2}

We simulate influence cascade in the four real-world networks, the GCTC, the SBM, the LFR 
and the ABCD networks.  We present results in this section.  

We simulate two most prevalent influence cascade models, the IC model 
and the LT model 
\cite{shakarian2015independent}.  In both models, the state of a vertex 
can be either active or inactive at any time step. Initially at time zero, a certain number
of vertices are selected to be active.  They are called the seeds of the diffusion.
In the independent cascade model, when an inactive vertex becomes activated, it 
will independently activate each of its currently inactive neighbors with probability 
$p$ in the next time step. Each active vertex has exactly one opportunity to influence its 
currently inactive neighbors. In our experiment, we set $p=0.15$.
In the linear threshold model, each inactive vertex, say $v$, 
computes the fraction of its active neighbors to its degree. The vertex switches to the active
state if the fraction exceeds a threshold $t$.  In the experiment we set $t=0.35$. 
In both IC and LT models, the influence diffusion process unfolds in discrete 
time steps until no more vertices can be activated.  We call the fraction of active vertices 
the influence spread of the diffusion process.  This quantity is shown in columns 
{\bf IC} and {\bf LT}.
From the entries in these two columns, we see that the GCTC model outperforms the other three 
models in predicting the influence spreads in both the IC model and the LT model in most cases.
The only except is the Email network.  In both the independent cascade model and the 
linear threshold model, the influence spreads of the four random network models are 
very close.  GCTC is not the worst among the four models.
Studies in \cite{morris2000,Watts.2007} show that densely connected clusters
affect significantly how opinions diffuse in a network.  Recall that the degree correlations
and transitivity of the GCTC model have the best match with those of real-world networks.
It seems to imply that the influence diffusion problem is quite sensitive to the
degree correlation and the clustering coefficient of the network in which the 
influence diffuses.  In \rfig{DBLP IC} and \rfig{DBLP LT} we present the influence 
spread in the DBLP network and the four random
network models as functions of $p$ and $t$, respectively.  
From these two figures, we see that the influence spread of 
the real network can be very different from those in random networks for 
certain range of $p$ and $t$. However, the curves corresponding to the GCTC model trace 
closely with those of the real network.

\bfig{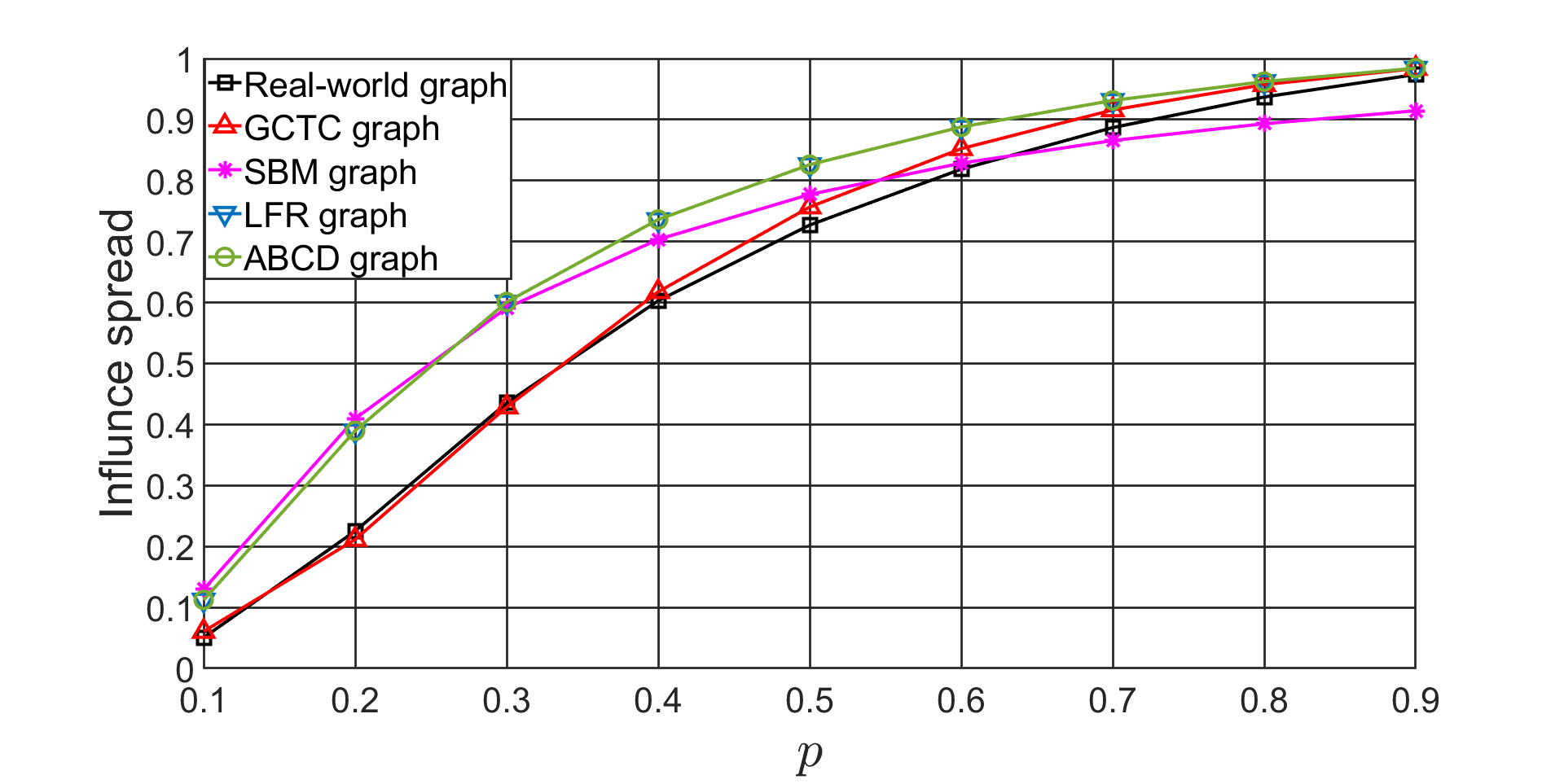}{3.8in}
\efig{DBLP IC}{Influence spread of the IC model on DBLP network}
\bfig{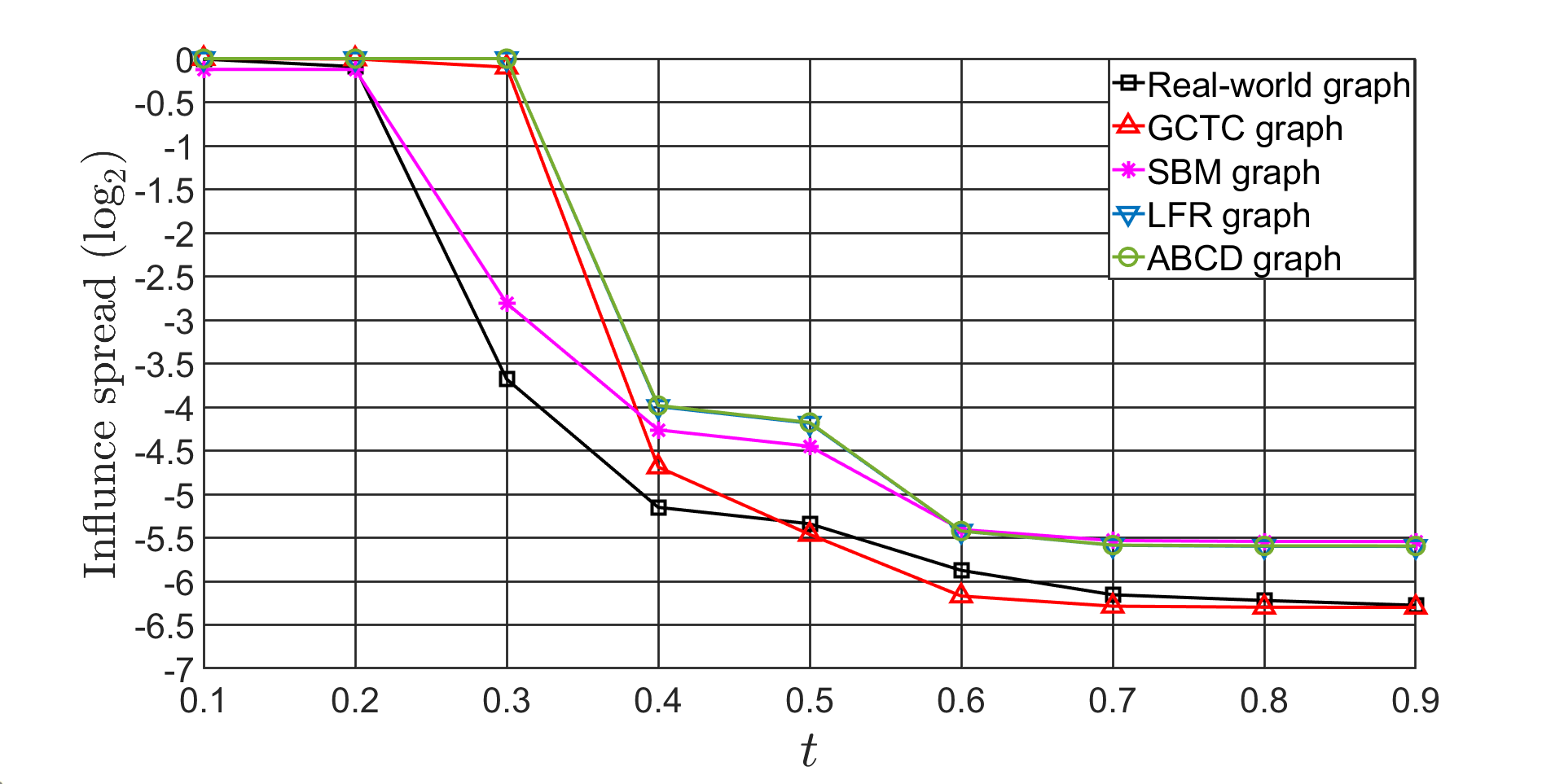}{3.8in}
\efig{DBLP LT}{Influence spread of the LT model on DBLP network}

\section{Conclusions}\label{s:conclusions}

In this paper, we have presented a generalized configuration model with triadic closure.
This model is an 
extension of the generalized configuration model by adding an additional layer of blocks
and random triadic closure operations.  The GCTC model  
possesses five most important properties of graphs that arise in network science.  
We have analyzed the 
Pearson degree correlation and clustering coefficient of the GCTC model. 
We have applied the GCTC model, the SBM, the LFR model and the ABCD model 
to model four networks collected in the real world.  We have shown that
by choosing parameters properly the GCTC model matches much better 
its clustering coefficient and Pearson degree
correlation coefficient with those of the data sets than the other three random models.
We have used the GCTC model, the SBM, the LFR model, the ABCD model and the four real-world
networks to examine three community detection algorithms and the
influence diffusion problem.  We found that the GCTC model, the SBM, the LFR model
and the ABCD model perform nearly equally well as a benchmark model
for community detection with GCTC model slightly better than the other three
models. We have simulated influence diffusion in the four models
and the four real-world networks using the independent cascade model
and the linear threshold model.  We found that the fraction of
nodes influenced in the GCTC model matches much better with that of
real-world networks than the other three models.

\centerline{Appendix A}

In this appendix we list all expectations needed to compute the covariance in Eq. (\ref{cov})
and the variance in Eq. (\ref{var}).  Note that $\ex[X]$ and $\ex[XY]$ have been derived in
\cite{Lee2019}.  We repeat them here for easy reference.

\begin{align}
\ex[X] &= \frac{\ex[Z^2]}{\ex[Z]}\label{X} \\
\ex[X^2] &= \frac{\ex[Z^3]}{\ex[Z]}\label{X^2} \\
\ex[XY] &=\frac{1-q}{(\ex[Z])^2}\sum_{i=1}^{b}\sum_{j=1}^{b}u_{i}u_{j}\nonumber\\
&+\frac{qb}{(\ex[Z])^2}\sum_{i=1}^{b}u_{i}u_{h(i)}\label{XY} \\
\ex[XY^2] &= \ex[X^2 Y]\nonumber\\
&= \ex[X^2\ex[Y|X]]\nonumber\\
&=\frac{(1-q)\ex[Z^2]\ex[Z^3]}{(\ex[Z])^2}\nonumber\\
&+\frac{qb}{(\ex[Z])^2}\sum_{i=1}^{b}u_{i}t_{h(i)}\label{XY^2} \\
\ex[X\ex[Y|X]] &= \ex[Y\ex[X | Y]]=\ex[XY] \label{XexY}\\
\ex[Y\ex[Y|X]] &= \ex[X\ex[X | Y]] \nonumber\\
&=\ex[(\ex[Y|X])^2]\nonumber\\
&=\frac{(1-q^2)(\ex[Z^2])^2}{(\ex[Z])^2}\nonumber\\ 
&+\frac{q^2b}{(\ex[Z])^2}\sum_{i=1}^{b}u_{i}u_{i} \label{YexY}\\
\ex[X(\ex[Y|X])^2] &=\frac{(1-q)^2(\ex[Z^2])^3}{(\ex[Z])^3}\nonumber\\
&+\frac{2(1-q)qb\ex[Z^2]}{(\ex[Z])^3}\sum_{i=1}^{b}u_{i}u_{h(i)}\nonumber\\
&+\frac{q^2b^2}{(\ex[Z])^3}\sum_{i=1}^{b}u_{i}u_{h(i)}u_{h(i)}\label{XexY2}\\
\ex[X^2(\ex[Y|X])^2]&= (1-q)^2\frac{(\ex[Z^2])^2\ex[Z^3]}{(\ex[Z])^3}\nonumber\\
&\quad+2(1-q)qb\frac{\ex[Z^2]}{(\ex[Z])^3}\sum_{i=1}^bu_it_{h(i)}\nonumber\\
&\quad+q^2b^2\frac{1}{(\ex[Z])^3}\sum_{i=1}^b(u_i)^2t_{h(i)}\label{X^2exY2}\\
\ex[XY\ex[Y|X]] &= \ex[XY\ex[X|Y]]\nonumber\\ 
&=\frac{(1-q)^2(\ex[Z^2])^3}{(\ex[Z])^3}\nonumber\\ 
&+\frac{2(1-q)qb\ex[Z^2]}{(\ex[Z])^3}\sum_{i=1}^{b}u_{i}u_{h(i)}\nonumber\\
&+\frac{q^2b^2}{(\ex[Z])^3}\sum_{i=1}^{b}u_{i}u_{h(i)}u_{h(i)}\label{XYexY}\\
\ex[\ex[Y|X]\ex[X|Y]] &= \frac{(1-q)(1+q+q^2)(\ex[Z^2])^2}{(\ex[Z])^2}\nonumber\\ 
&+\frac{q^3b}{(\ex[Z])^2}\sum_{i=1}^{b}u_{i}u_{h(i)}\label{exYexX}\\ 
\ex[Y\ex[Y|X]\ex[X|Y]] &= \frac{(1+q)(1-q)^2(\ex[Z^2])^3}{(\ex[Z])^3}\nonumber\\ 
&+\frac{(1-q^2)qb\ex[Z^2]}{(\ex[Z])^3}\sum_{i=1}^{b}u_{i}u_{h(i)}\nonumber\\ 
&+\frac{(1-q)q^2b\ex[Z^2]}{(\ex[Z])^3}\sum_{i=1}^{b}u_{i}u_{i}\nonumber\\ 
&+\frac{q^3b^2}{(\ex[Z])^3}\sum_{i=1}^{b}u_{i}u_{i}u_{h(i)}\label{YexYexX}\\ 
\ex[XY\ex[Y|X]\ex[X|Y]] &=  \frac{(1-q)^3(\ex[Z^2])^4}{(\ex[Z])^4}\nonumber\\ 
&+\frac{3qb(1-q)^2(\ex[Z^2])^2}{(\ex[Z])^4}\sum_{i=1}^{b}u_{i}u_{h(i)}\nonumber\\ 
&+\frac{q^2b^2(1-q)}{(\ex[Z])^4}(\sum_{i=1}^{b}u_{i}u_{h(i)})^2\nonumber\\ 
&+\frac{2q^2b^2(1-q)\ex[Z^2]}{(\ex[Z])^4}\sum_{i=1}^{b}u_{i}u_{i}u_{h(i)}\nonumber\\
&+\frac{q^3b^3}{(\ex[Z])^4}\sum_{i=1}^{b}u_{i}u_{i}u_{h(i)}u_{h(i)}\label{XYexYexX}
\end{align}

\centerline{Appendix B}
In Appendix B, we present the proof of \rthe{thm1}. 
Recall that stubs corresponding to the degrees are partitioned {\em evenly} into
$b$ blocks.  Recall also that our construction algorithm arranges degrees in ascending order (descending order
will also work).  That is,  
\[
x\le y\ \mbox{for all}\ x\in H_i\ \mbox{and}\ y\in H_j,
\] 
where $i\le j$.
Due to \rassum{1}, \req{evenblock} holds.
We break down the proof of \rthe{thm1} into several lemmas listed as follows.

\blem{lem-main}
Suppose that $h(i)=i$.   If sequences $\{x_i, i=1, 2, \ldots, b\}$ and 
$\{y_i, i=1, 2, \ldots, b\}$ are both
non-decreasing or both non-increasing, then
\begin{equation}\label{main-ineq}
b\sum_{i=1}^{b}x_{i}y_{h(i)}-\sum_{i=1}^{b}x_{i}\sum_{j=1}^{b}y_{j}\ge 0.
\end{equation}
\elem
\bproof{of \rlem{lem-main}.} 
If sequences $\{x_i\}$ and $\{y_i\}$ are both non-increasing, then
\begin{align*}
x_{[i]}&= x_i \\
y_{[i]}&= y_i,
\end{align*}
where $x_{[i]}$ denotes the $i$-th largest element in sequence $\{x_i\}$.
On the other hand, if sequences $\{x_i\}$ and $\{y_i\}$ are both non-decreasing, then
\begin{align*}
x_{[i]}&= x_{b-i+1} \\
y_{[i]}&= y_{b-i+1}.
\end{align*}
In either cases, we have
\beq{main0000}
\sum_{i=1}^b x_i y_{h(i)}=\sum_{i=1}^b x_i y_i = \sum_{i=1}^b x_{[i]} y_{[i]}.
\eeq
Now consider circular shift permutation $\sigma_j(\cdot)$ with $\sigma_j(i)=(i+j-1\ \mbox{mod}
\ b)+1$ for $j=1, 2, \ldots, b$.  From symmetry, we have $\sigma_j(i)=\sigma_i(j)$.  Thus,
\beq{main1111}
\sum_{i=1}^b \sum_{j=1}^b x_i y_j=\sum_{i=1}^b \sum_{j=1}^b x_i y_{\sigma_i(j)}=
\sum_{j=1}^b \sum_{i=1}^b x_i y_{\sigma_j(i)}.
\eeq
Let
\[
v_i=y_{\sigma_j(i)}.
\]
Thus,
\beq{main2222}
\sum_{i=1}^b x_i y_{\sigma_j(i)} = \sum_{i=1}^b x_i v_i
\le \sum_{i=1}^b x_{[i]} v_{[i]}.
\eeq
The last inequality in \req{main2222} is due to the well known 
Hardy, Littlewood and P\'olya  rearrangement inequality
(see e.g., the book \cite{Marshall.11}, pp. 141).  
Clearly, sequence $\{v_i\}$ is a shifted version of $\{y_i\}$.  Thus,
\[
v_{[i]}=y_{[i]}.
\]
Substituting the preceding equation and \req{main2222} into \req{main1111}, we obtain
\[
\sum_{i=1}^b \sum_{j=1}^b x_i y_j \le
\sum_{j=1}^b \sum_{i=1}^b x_{[i]} y_{[i]}
=b \sum_{i=1}^ b x_i y_i,
\] 
where the last equality in the preceding is due to \req{main0000}.
\eproof

\blem{lem-ascend}
If stubs corresponding to degrees are arranged in ascending order evenly into blocks, then
\begin{enumerate}
	\item $\{u_i: 1\le i\le b\}$
	\item $\{t_i: 1\le i\le b\}$
	\item $\{t_i-u_i: 1\le i\le b\}$
	\item $\{u_i^2: 1\le i\le b\}$
	\item $\{u_i(u_i-c): 1\le i\le b\}$ for any constant $c$
\end{enumerate}
are all non-deceasing sequences.
\elem
\bproof{of \rlem{lem-ascend}.} Let $i$ and $j$ be two blocks, where
$i<j$.  From \rassum{1},
\beq{even-blocks}
\sum_{x\in H_i} x p_x = \sum_{x\in H_j} x p_x.
\eeq
Let $x_i^{\mbox{\small max}}$ denote the maximum degree in block $i$.  Then,
\begin{align*}
u_i &= \sum_{x\in H_i} x^2 p_x \\
&\le x_i^{\mbox{\small max}} \sum_{x\in H_i} x p_x \\
&=x_i^{\mbox{\small max}} \sum_{x\in H_j} x p_x \\
&\le \sum_{x\in H_j} x^2 p_x \\
&=u_j.
\end{align*}
Other sequences can be proved similarly.
\eproof

\blem{lem1}
If $h(i)=i$, then
\[
W_i \ge 0
\]
for $i=1, 2, 3, 4, 5$.
\elem
\bproof{of \rlem{lem1}.} 
Lee \etal \cite{Lee2019} proved that $W_1\ge 0$ if $h(i)=i$.  We now prove
that $W_2 \ge 0$.  From part 2 of \rlem{lem-ascend}, it follows that
sequence $\{t_i: 1\le i\le b\}$ is non-decreasing.  It follows from \rlem{lem-main}
that $W_2\ge 0$.  Proof for $W_i\ge 0$, $3\le i\le 5$ is similar.
\eproof

\blem{lem2}
If $h(i)=i$, then
\[
W_2 \ge W_1.
\]
\elem
\bproof{of \rlem{lem2}.} One can express
\begin{align*}
W_2 - W_1 &= b\sum_{i=1}^b u_i(t_{h(i)}-u_{h(i)})-\sum_{i=1}^b\sum_{j=1}^b u_i(t_j-u_j)\\
&=b\sum_{i=1}^b u_i(t_{i}-u_{i})-\sum_{i=1}^b\sum_{j=1}^b u_i(t_j-u_j).
\end{align*}
From part 3 of \rlem{lem-ascend}, it follows that sequence $\{t_i-u_i: 1\le i\le b\}$
is non-decreasing.  The claim of the lemma follows from \rlem{lem-main}.
\eproof

\blem{lem3}
If $h(i)=i$, then
\begin{align}
&\ex[Z]\ex[Z^3]-(\ex[Z^{2}])^{2} \ge W_{1} \label{lem3-1}\\
&W_4-\frac{\ex[Z]}{b} W_3 \ge 0 \label{lem3-2}
\end{align}
\elem
\bproof{of \rlem{lem3}.}  We first prove (\ref{lem3-1}).  Note that
\begin{align*}
&\ex[Z^3]\ex[Z]-(\ex[Z^2])^2-W_1 \\
&= bz\sum_{i=1}^b t_i -(\sum_{i=1}^b u_i)^2-(b\sum_{i=1}^b u_i^2-
\sum_{i=1}^b u_i \sum_{j=1}^b u_j) \\
&= b\sum_{i=1}^b (t_i z-u_i^2),
\end{align*}
We claim that $t_i z \ge u_i^2$ for all $i$.  Inequality (\ref{lem3-1}) follows directly
from this claim.  We now prove the claim.  From the definition of $u_i$ and $t_i$
in (\ref{def-u}) and (\ref{def-t}) respectively, we have
\begin{align*}
&t_i z-u_i^2 \\
&=\sum_{x\in H_i}\sum_{y\in H_i} x^3 y p_x p_y - \sum_{x\in H_i}\sum_{y\in H_i} x^2 y^2 p_x p_y \\
&=\sum_{x\in H_i}\sum_{y\in H_i} x^2 y (x-y)p_x p_y \\
&=\sum_{x, y\in H_i, x> y} x^2 y (x-y)p_x p_y + \sum_{x, y\in H_i, x < y} x^2 y (x-y)p_x p_y \\
&=\sum_{x, y\in H_i, x> y} x^2 y (x-y)p_x p_y - \sum_{x, y\in H_i, x < y} x^2 y (y-x)p_x p_y\\
&=\sum_{x, y\in H_i, x> y} x^2 y (x-y)p_x p_y - \sum_{x, y\in H_i, y < x} y^2 x (x-y)p_y p_x,
\end{align*}
where the last equality follows by exchanging symbols $x$ and $y$ in the second term
of the last equation.  The preceding difference equals
\[
\sum_{x, y\in H_i, x> y} x y (x-y)^2 p_x p_y,
\]
which is non-negative.

Next we prove (\ref{lem3-2}).
Note that $\frac{\ex[Z]}{b}=z \le u_{i}$ for $i=1,2,...,b$.
Substituting (\ref{W_3}) and (\ref{W_4})  
into  (\ref{lem3-2}), we obtain
\begin{align} 
&W_{4}-\frac{\ex[Z]}{b}W_{3}\nonumber \\
&=\left(b\sum_{i=1}^{b}u_{i}u_{i}u_{i}-\sum_{i=1}^{b}u_{i}u_{i}\sum_{j=1}^{b}u_{j}\right)\nonumber \\
&\quad-\left(b\sum_{i=1}^{b}u_{i}u_{i}z-\sum_{i=1}^{b}u_{i}z\sum_{j=1}^{b}u_{j}\right)\nonumber \\
& =b\sum_{i=1}^{b}u_{i}u_{i}(u_{i}-z)-\sum_{i=1}^{b}u_{i}\sum_{j=1}^{b}u_{j}(u_{j}-z).\nonumber
\end{align}
From part 3 of \rlem{lem-ascend}, it follows that sequence $\{u_i(u_i-z): 1\le i\le b\}$
is non-decreasing.  Inequality (\ref{lem3-2}) follows from \rlem{lem-main}.
\eproof

Now we prove \rthe{thm1}.\newline
\bproof{of \rthe{thm1}.}   Through extensive algebraic manipulation, we express the sum of the first three terms and the sum
of the last three terms in (\ref{cov-final}) in a different manner, {\em i.e.}
\begin{align}
&\alpha_{0} + \beta_{1}W_{1}+ \beta_{2}W_{2}
= D_{1}+D_{2}+D_{3}+D_{4}+D_{5} \\
&\beta_{3}W_{3}+ \beta_{4}W_{4}+ \beta_{5}W_{5} = D_{6} + D_{7},
\end{align}
where
\begin{align}
&D_{1} = \frac{q}{(\ex[Z])^2}W_{1}=\cov(X,Y)\label{D1} \\
&D_{2}=2a^{2}\frac{\ex[Z^2]}{(\ex[Z])^{3}}\left(\ex[Z]\ex[Z^3]-(\ex[Z^{2}])^{2}\right) 
\nonumber \\
&\quad - 2a^{2}q\frac{\ex[Z^2]}{(\ex[Z])^3} W_{1}\nonumber \\
&\quad+2a\frac{q(1-q)\ex[Z^2]}{(\ex[Z])^3} W_{1}\nonumber \\
&\quad-2a^{2}\frac{q(1-q)\ex[Z^{2}]}{(\ex[Z])^{3}}W_{1}\label{D2}\\
&D_{3}=2a(1-a)\frac{1}{(\ex[Z])^{2}}\left(\ex[Z]\ex[Z^3]-(\ex[Z^{2}])^{2}\right) \nonumber \\
&\quad+2a^{2}\frac{q}{(\ex[Z])^{2}}W_{2} - 2a\frac{q}{(\ex[Z])^2} W_{1}\label{D3} \\
&D_{4}=a^{2}\frac{q(1-q)^{2}(\ex[Z^{2}])^{2}}{(\ex[Z])^{4}}W_{1}\label{D4} \\
&D_{5}=a^{2}\frac{q^{3}}{(\ex[Z])^{2}}W_{1} \label{D5}\\
&D_6=2a \frac{q^{2}b}{(\ex[Z])^{3}}\left(1-a+\frac{(1-q)(\ex[Z^{2}])^{2}}{\ex[Z]}\right) \nonumber\\
&\quad \times \left(W_4-\frac{\ex[Z]}{b}W_3\right)\label{D6} \\
& D_{7} = a^{2}\frac{q^{3}b^{2}}{(\ex[Z])^{4}}\left(W_{5} -2\frac{\ex[Z]}{b}W_{4}\right).
\label{D7} 
\end{align}

We claim that 
\[
D_i\ge 0, \quad i=1, 2, 3, \ldots, 7.
\]
Since $D_1=\cov(X,Y)$, the claim implies (\ref{thm1-1}).  In the rest of the proof, we
focus on the proof of the claim.

From \rlem{lem1}, $W_1\ge 0$. It follows from (\ref{D4}) and (\ref{D5}) that
$D_4\ge 0$ and $D_5\ge 0$.  For $D_2$, we place $\ex[Z]\ex[Z^3]-(\ex[Z^{2}])^{2}$ with
$W_1$ and obtain a lower bound for $D_2$, {\em i.e.}
\[
D_2 \ge 2a(1-q)(a+(1-a)q)\frac{\ex[Z^2]}{(\ex[Z])^3} W_{1},
\]
which is greater than or equal to zero, because $W_1\ge 0$.  

Now we consider $D_3$. We replace $\ex[Z]\ex[Z^3]-(\ex[Z^{2}])^{2}$ and $W_2$ with
$W_1$ and obtain a lower bound for $D_3$, {\em i.e.}
\[
D_{3} \ge 2a(1-a)(1-q)\frac{1}{(\ex[Z])^2} W_{1},
\]
which is greater than or equal to zero, again because $W_1\ge 0$.

Now we analyze  $D_6$.  From (\ref{D6}) it is clear that $D_6\ge 0$, if and only if 
\[
W_4-\frac{\ex[Z]}{b} W_3 \ge 0.
\]
From (\ref{lem3-2}) in \rlem{lem3}, $W_4-\frac{\ex[Z]}{b} W_3 \ge 0$ holds. Then, we have $D_6\ge 0$.

Finally we analyze $D_{7}$.  From (\ref{D7}), $D_7\ge 0$, if and only if
\[
W_{5}-2\frac{\ex[Z]}{b}W_{4}\ge 0.
\]
Replacing $W_4$ and $W_5$ with their definitions in (\ref{W_4}) and (\ref{W_5}), we have
\begin{align} 
 &W_{5}-2\frac{\ex[Z]}{b}W_{4}\nonumber \\
 &=\left(b\sum_{i=1}^{b}u_{i}u_{i}u_{i}u_{i}-\sum_{i=1}^{b}u_{i}u_{i}\sum_{j=1}^{b}u_{j}u_{j}\right)\nonumber \\
 &\qquad-2\left(b\sum_{i=1}^{b}u_{i}u_{i}u_{i}z-\sum_{i=1}^{b}u_{i}u_{i}\sum_{j=1}^{b}u_{j}z\right )\nonumber \\
&=b\sum_{i=1}^{b}u_{i}u_{i}u_{i}(u_{i}-2z)-\sum_{i=1}^{b}u_{i}u_{i}\sum_{j=1}^{b}u_{j}(u_{j}-2z).\label{D7_1}
\end{align}
From case (4) and case (5) of \rlem{lem-ascend} and \rlem{lem-main}, it follows that the right
side of (\ref{D7_1}) is non-negative.  The proof of \rthe{thm1} is completed.

\eproof

\centerline{Appendix C}

In Appendix C, we analyze $\sigma_{X+X'}$ in the denominator in Eq. (\ref{rho}). 
Among the expectation terms needed in the $\sigma_{X+X'}$, 
$\ex[X]$ and $\ex[X']$ were already analyzed. We next analyze $\ex[X X']$ and $\ex[(X')^2]$.
Since $Y_i$ and $Y$ are identically distributed and $\ex[Y_i | X]=\ex[Y|X]$. From (\ref{nte}) we have
\begin{align}
&\ex[XX'] \nonumber\\
&= \ex[\ex[XX' | X, Y, Y_i,\forall i]] \nonumber\\
&= \ex\left[\ex\left[X \left(\sum_{j=1}^{Y-1}B_{1,j}+\sum_{i=2}^X \sum_{j=1}^{Y_i-1} B_{ij}
\right)\Biggl| X, Y, Y_i,\forall i\right]\right]\nonumber\\
&= \ex\left[\ex\left[X \left(a(Y-1)+a\sum_{i=2}^X (Y_i-1)
\right)\Biggl| X, Y, Y_i,\forall i\right]\right] \nonumber\\
&= \ex\Big[a X(\ex[Y|X]-1+(X-1)(\ex[Y|X]-1))\Big]\nonumber\\
&= a\left(\ex[X^2 Y]-\ex[X^2]\right).\label{XX'}
\end{align}

Next, we analyze $\ex[(X')^2]$.
From (\ref{nte}), we have
\begin{align}
&\ex[(X')^2]\nonumber\\
&=\ex\left[\ex\left[\left(\sum_{j=1}^{Y-1}B_{1,j}+\sum_{i=2}^X \sum_{j=1}^{Y_i-1} B_{ij}
\right)^2\Biggl|X, Y, Y_i, \forall i\right]\right]\nonumber\\
&= \ex\left[\ex\left[\left(\sum_{j=1}^{Y-1}B_{1,j}\right)^2
\Biggl|X, Y, Y_i, \forall i\right]\right] \nonumber\\
&\ +\ex\left[\ex\left[2\left(\sum_{j=1}^{Y-1}B_{1,j}\right)
\left(\sum_{i=2}^X \sum_{j=1}^{Y_i-1} B_{ij}\right)\Biggl|X, Y, Y_i, \forall i\right]\right] 
\nonumber\\
&\ + \ex\left[\ex\left[\left(\sum_{i=2}^X \sum_{j=1}^{Y_i-1} B_{ij}\right)^2
\Biggl|X, Y, Y_i, \forall i\right]\right].\label{X'sq}
\end{align}
The first term on the right side of Eq. (\ref{X'sq}) is equal to
\begin{align*}
& \ex\left[\ex\left[\sum_{i=1}^{Y-1}\sum_{j=1}^{Y-1}B_{1,i}B_{1,j}
\Biggl|X, Y, Y_i, \forall i\right]\right] \\
&= \ex\left[\ex\left[\sum_{i=1}^{Y-1}B_{1,i}^2+
\sum_{i=1}^{Y-1}\sum_{\twoLineSub{j=1}{j\ne i}}^{Y-1}B_{1,i}B_{1,j}
\Biggl|X, Y, Y_i, \forall i\right]\right] \\
&=a(\ex[Y]-1)+a^2 \ex[(\ex[Y | X]-1)(\ex[Y | X]-2)]\\
&=a^2\ex[(\ex[Y | X])^2]+(a-3a^2)\ex[Y]+2a^2-a.
\end{align*}
The second term on the right side of Eq. (\ref{X'sq}) is equal to
\begin{align*}
&\ex\left[\ex\left[2\left(\sum_{j=1}^{Y-1}B_{1,j}\right)
\left(\sum_{i=2}^X \sum_{j=1}^{Y_i-1} B_{ij}\right)\Biggl|X, Y, Y_i, \forall i\right]\right]  \\
&=\ex\left[\ex\left[2(Y-1)\cdot a^2\cdot \left(\sum_{i=2}^X Y_i - (X-1)\right)\Biggl|X, Y, Y_i, \forall i\right]\right].
\end{align*}
Taking average of the preceding with respect to $Y_i$ for all $i$, while conditioning
on $X$ and $Y$, we have
\begin{align*}
&\ex\left[\ex\left[2\left(\sum_{j=1}^{Y-1}B_{1,j}\right)
\left(\sum_{i=2}^X \sum_{j=1}^{Y_i-1} B_{ij}\right)\Biggl|X, Y, Y_i, \forall i\right]\right]  \\
&=2a^2\ex\left[(Y-1)(X-1)(\ex[Y | X]-1)\right]\\
&=2a^2(-2\ex[XY]+3\ex[X]+\ex[XY\ex[Y | X]]\\
&\quad-\ex[Y\ex[Y | X]]-1).
\end{align*}
Now consider the third term on the right of (\ref{X'sq}).
It can be written as
\begin{align}
&\left(\sum_{i=2}^X \sum_{j=1}^{Y_i-1} B_{ij}\right)^2\nonumber\\
&= \sum_{i=2}^X \sum_{j=1}^{Y_i -1} B_{i,j}^2 +
\sum_{i=2}^X \sum_{j=1}^{Y_i-1} \sum_{\twoLineSub{k=2}{k\ne i}}^X\sum_{\ell=1}^{Y_k-1}
B_{i,j}B_{k,\ell} \nonumber\\
&\quad +\sum_{i=2}^X\sum_{j=1}^{Y_i-1}\sum_{\twoLineSub{\ell=1}{\ell\ne j}}^{Y_i-1} B_{i,j}
B_{i,\ell}.\label{3-1}
\end{align}
The conditional expectation of (\ref{3-1}), given $X$, $Y$ and $Y_i$ for all $i$, is
\begin{align}
&a\sum_{i=2}^X (Y_i-1)+a^2\sum_{i=2}^X\sum_{\twoLineSub{k=2}{k\ne i}}^X
(Y_i-1)(Y_k-1) \nonumber\\
&\ \ +a^2\sum_{i=2}^X(Y_i-1)(Y_i-2).\label{3-2}
\end{align}
Taking average on (\ref{3-2}) with respect to $Y_i$ for all $i$, we have
\begin{align*}
&a(X-1)(\ex[Y | X]-1)\nonumber\\
&+a^2(X-1)(X-2)\ex[(Y_2-1)(Y_3-1)] | X)) \nonumber\\
&+a^2 (X-1)\ex[(Y-1)(Y-2) | X].
\end{align*}
Since $Y_2$ and $Y_3$ are conditional independent, given $X$, the preceding quantity is
equal to
\begin{align}
&a(X-1)(\ex[Y | X]-1)\nonumber\\
&+a^2(X-1)(X-2)(\ex[Y| X-1 ])^2) \nonumber\\
&+a^2 (X-1)\ex[(Y-1)(Y-2) | X].\label{3-3}
\end{align}
Finally, taking average on (\ref{3-3}) with respect to $X$, we obtain
\begin{align}
&a(\ex[XY]-2\ex[X]+1)+\nonumber\\
&a^2(-\ex[X^2Y]+3\ex[XY] -2\ex[Y]\nonumber\\
&\quad+\ex[X^2\ex[Y | X]\ex[Y | X]]\nonumber\\
&\quad-3\ex[X\ex[Y | X]\ex[Y | X]]\nonumber\\
&\quad+2\ex[\ex[Y | X]\ex[Y | X]]).\label{3-final}
\end{align}
This is the third term on the right side of (\ref{X'sq}).

\bibliography{bibdatabase.bib}
\bibliographystyle{ieeetran}

\end{document}